\documentclass[reprint,aps,amsmath,amssymb,twoside,prd,showkeys,superscriptaddress]{revtex4-1}
\pdfoutput=1
\usepackage{verbatim}
\usepackage{comment}
\usepackage{graphicx}
\usepackage[T1]{fontenc}
\usepackage{graphicx}
\usepackage{epsfig}
\usepackage{bm}
\usepackage{tensor}
\usepackage{amsfonts}
\usepackage{amssymb}
\usepackage{float}
\usepackage{amsmath}
\usepackage{subfigure}
\usepackage{dcolumn}
\usepackage{cancel}
\usepackage[colorlinks]{hyperref}
\usepackage[utf8]{inputenc}
\usepackage[usenames,dvipsnames]{color}
\hypersetup{
     breaklinks=true,
    pdfstartview={FitH},    
    colorlinks=true,       
    linkcolor=blue,          
    citecolor=red,        
    filecolor=magenta,      
    urlcolor=blue,           
    anchorcolor=green,      
    linktocpage=true
}

\usepackage{amssymb,amsmath}
\usepackage{graphicx}
\usepackage{color}
\usepackage{dcolumn}
\usepackage{pgfplots}
\usepackage{booktabs}
\usepackage{multirow}
\usepackage{dcolumn}
\usepackage{amsmath}
\usepackage{mathtools}
\usepackage{amsfonts}
\usepackage{amssymb}
\usepackage{epstopdf}
\usepackage{bm}
\usepackage{siunitx}
\usepackage{braket}
\usepackage{enumitem}
\usepackage{soul}
\usepackage{ulem}
\usepackage{color}
\usepackage{transparent}
\usepackage{pifont}
\usepackage[font={small}]{caption}
\newcommand\rf[1]{(\ref{eq:#1})}
\newcommand\lab[1]{\label{eq:#1}}
\newcommand\nonu{\nonumber}
\newcommand\br{\begin{eqnarray}}
\newcommand\er{\end{eqnarray}}
\newcommand\be{\begin{equation}}
\newcommand\ee{\end{equation}}

\newcommand\lb{\lbrack}
\newcommand\rb{\rbrack}

\renewcommand\({\left(}
\renewcommand\){\right)}

\newcommand\bc{\begin{center}}
\newcommand\ec{\end{center}}

\newcommand\partder[2]{\frac{{\partial {#1}}}{{\partial {#2}}}}

\renewcommand\a{\alpha}

\renewcommand\d{\delta}

\newcommand\eps{\epsilon}
\newcommand\vareps{\varepsilon}

\newcommand\G{\Gamma}

\newcommand\h{\frac{1}{2}}
\renewcommand\k{\kappa}
\renewcommand\l{\lambda}

\newcommand\m{\mu}
\newcommand\n{\nu}

\newcommand\p{\phi}
\newcommand\vp{\varphi}
\renewcommand\P{\Phi}
\newcommand\pa{\partial}

\renewcommand\th{\theta}

\newcommand\cP{{\mathcal P}}

\newcommand\vpdot{\stackrel{.}{\varphi}}
\newcommand\vpddot{\stackrel{..}{\varphi}}

\newcommand\adot{\stackrel{.}{a}}
\newcommand\addot{\stackrel{..}{a}}
\newcommand\Hdot{\stackrel{.}{H}}

\newcommand{\afffias}{Frankfurt Institute for Advanced Studies (FIAS), Ruth-Moufang-Strasse~1, 60438 Frankfurt am Main, Germany}
\newcommand{\affbgu}{Physics Department, Ben-Gurion University of the Negev, Beer-Sheva 84105, Israel}
\newcommand{\affcam}{DAMTP, Centre for Mathematical Sciences, University of Cambridge, Wilberforce Road, Cambridge CB3 0WA, UK}

\newcommand{\affckavli}{Kavli Institute of Cosmology (KICC), University of Cambridge, Madingley Road, Cambridge, CB3 0HA, UK}
\newcommand{\affbaham}{Bahamas Advanced Study Institute and Conferences, 4A Ocean Heights, Hill View Circle, Stella Maris, Long Island, Bahamas.}
\newcommand{\affchil}{Instituto de F\'{\i}sica, Pontificia Universidad Cat\'{o}lica de Valpara\'{\i}so, Avenida Brasil 2950, Casilla 4059, Valpara\'{\i}so, Chile.}
\newcommand{\affif}{IFPU – Institute for Fundamental Physics of the Universe, via Beirut 2, 34151, Trieste, Italy}

\begin{document}

\title{Unifying Inflation  {with early and late Dark Energy} in Multi-Fields:\\ Spontaneously broken scale invariant TMT}

\author{Eduardo Guendelman}
\email{guendel@bgu.ac.il}
\affiliation{\affbgu}\affiliation{\afffias}\affiliation{\affbaham}\affiliation{\affif}
\author{Ram\'{o}n Herrera}
\email{ramon.herrera@pucv.cl}
\affiliation{\affchil}
\author{David Benisty}
\email{db888@cam.ac.uk}
\affiliation{\affcam}\affiliation{\affckavli}\affiliation{\affif}
\begin{abstract}
 {A unified multi scalar field model with three flat regions is discussed. The three flat regions are the inflation, early and late dark energy epochs. The potential is obtained by a spontaneous breaking of scale invariance generated by Non Riemannian Measures of integration (or Two Measures Theories (TMT)).}   
We define the scale invariant couplings of the scalar fields to the different measures through exponential potentials. Spontaneous breaking of scale invariance takes place when integrating the fields that define the measures. When going to the Einstein frame we obtain: (i) An
effective potential for the scalar fields with three flat regions which allows for a unified description of both early universe inflation (in the higher energy density flat region) as well as of present dark energy epoch which can be realized with a double phase, i.e., in two flat regions. (ii) In the slow roll inflation, only one field combination
the ``dilaton", which transforms under scale transformations, has non trivial dynamics, the orthogonal one, which is scale invariant  remains constant. 
The corresponding perturbations of the dilaton are calculated. 
(iii) For a reasonable choice of the parameters the present model perturbations conforms to the
Planck Collaboration data.
 (iv) In the late universe we define scale invariant couplings of Dark Matter to the dilaton. 
These couplings define a matter induced potential for the dilaton and extremizing this potential determines the scale invariant scalar field, while all
exotic non canonical behavior of the Dark Matter as well as any possible $5^{th}$ force disappear. (v) We calculate the evolution of the late universe under these conditions
with the realization of two different possible realizations of $\Lambda$CDM type scenarios depending of the flat region in the late universe. These two phases could appear at different times in the history of the universe.(vi) From the Planck data, we find the constraints on the parameters during the inflationary epoch and these values are used to obtain constraints relevant to the present epoch. 
\end{abstract}
\maketitle
\section{Introduction}
\label{intro}

In the ``standard cosmological''
framework for the early universe \cite{Kolb:1990vq,Mukhanov:2005sc,Guth:1980zm,Guth:1982ec,Starobinsky:1979ty,Starobinsky:1980te,Linde:1981mu,Albrecht:1982wi,Mukhanov:1981xt}
the universe starts with a period of exponential expansion called ``inflation''. 
At the same time, after the discovery of the accelerating universe  \cite{Bahcall:1999xn,Peebles:2002gy,SupernovaSearchTeam:1998fmf,SupernovaCosmologyProject:1998vns}, we have now a late universe ``standard cosmological''
framework for the late universe, the $\Lambda$CDM  picture \cite{Cen:1993az,Weinberg:1988cp}, consisting of a cosmological constant, Dark matter and ordinary visible matter, the Universe being now dominated by the Cosmological Constant or Dark Energy (DE) and the Dark Matter (DM). This simple $\Lambda$CDM is now being somewhat challenged by the discovery of several cosmological tensions, the $H_0$  tension \cite{Bernal:2016gxb,Graef:2018fzu,Bernal:2021yli,DiValentino:2021izs} and the $\sigma_8$ tension \cite{Keeley:2019esp,Pandey:2019plg,Abdughani:2019wss,Lambiase:2018ows,Lin:2019htv,Berbig:2020wve,Benisty:2020kdt}. This suggests that the introduction of a cosmological term to describe the DE and the addition of DM may be a too simple description of the late Universe.
In the inflationary period also primordial density perturbations are generated \cite{Mukhanov:2005sc}. The ``inflation'' is followed by particle creation, where the observed matter and radiation were generated, and finally the evolution arrives to a present phase of slowly accelerating universe. In this standard model, however, at least two fundamental questions remain unanswered:  

\begin{itemize}
\item 
The early inflation, although solving many cosmological puzzles,
like the horizon and  flatness problems, cannot address the initial singularity 
problem; 
\item 
There is no explanation for the existence of two periods of 
exponential expansion with such wildly different scales -- the inflationary 
phase and the present phase of slowly accelerated expansion of the universe.
\end{itemize}

The best known mechanism for generating a period of accelerated expansion 
is through the presence of some vacuum energy. In the context of a 
scalar field theory, vacuum energy density appears naturally when the scalar
field acquires an effective potential $U_{\rm eff}$ which has flat regions so 
that the scalar field can ``slowly roll'' \cite{Linde:1981mu,Albrecht:1982wi,Liddle:1992wi,Liddle:1993fq} and its 
kinetic energy can be neglected resulting in an energy-momentum tensor 
$T_{\m\n} \simeq - g_{\m\n} U_{\rm eff}$.

The possibility of continuously connecting an inflationary phase to a slowly 
accelerating universe through the evolution of a single scalar field -- the
{\em quintessential inflation scenario} -- has been first studied in 
Ref.\cite{Peebles:1998qn}. Also, 
$F(R)$ models can yield 
both an early time inflationary epoch and a late time de Sitter phase with 
vastly different values of effective vacuum energies \cite{Nojiri:2003ft,Cognola:2007zu,Appleby:2009uf}.
For a recent proposal of a quintessential inflation mechanism based on 
the k-essence  framework, see Ref.\cite{Saitou:2011hv}. For
another recent approach to quintessential inflation based on the 
``variable gravity'' model \cite{Wetterich:2013aca,Wetterich:2013jsa} and for extensive list of references 
to earlier work on the topic, see Ref.\cite{Hossain:2014xha}. Other ideas based on the so called $\alpha$ attractors \cite{Dimopoulos:2017zvq,Dimopoulos:2017tud,Linde:1981mu,Akrami:2017cir,Akrami:2020zxw,Rodrigues:2020jsv,Elizalde:2015nya,Dubinin:2017irq,Pozdeeva:2020shl,Herrera:2018cgi,Herrera:2018mvo,Herrera:2019xhs,Herrera:2020mjh,AresteSalo:2021wgb,AresteSalo:2021lmp,Gonzalez-Espinoza:2021qnv}, which uses non canonical kinetic terms have been studied. Finally a quintessential inflation based on a Lorentzian Slow Roll ansatz \cite{Benisty:2020qta,Benisty:2020qta} which automatically gives two flat regions.

In previous papers \cite{Guendelman:2014bva,Guendelman:2015liz,Guendelman:2014waa} we have studied a unified scenario where both an inflation 
and a slowly accelerated phase for the universe can appear naturally from the 
existence of two flat regions in the effective scalar field potential which
we derive systematically from a Lagrangian action principle. 
Namely, we started with a new kind of globally Weyl-scale invariant gravity-matter 
action within the first-order (Palatini) approach formulated in terms of two 
different non-Riemannian volume forms (integration measures) \cite{Guendelman:2014waa}.
In this new theory there is a single scalar field with kinetic terms coupled to 
both non-Riemannian measures, and in addition to the scalar curvature 
term $R$ also an $R^2$ term is included (which is similarly allowed by global 
Weyl-scale invariance). Scale invariance is spontaneously broken upon solving part 
of the corresponding equations of motion due to the appearance of two 
arbitrary dimensionful integration constants.


Let us briefly recall the origin of current approach. The main idea comes from 
Refs.\cite{Guendelman:1999qt,Guendelman:1999tb,Guendelman:2002js},
where some of us have proposed a new class of gravity-matter theories based on the 
idea that the action integral may contain a new metric-independent generally-covariant 
integration measure density, \textsl{i.e.}, an alternative non-Riemannian volume form 
on the space-time manifold defined in terms of an auxiliary antisymmetric gauge
field of maximal rank. The originally proposed modified-measure gravity-matter theories
\cite{Guendelman:1999qt,Guendelman:1999tb,Guendelman:2002js} contained two terms in the pertinent Lagrangian action
-- one with a non-Riemannian integration measure and a second one with the
standard Riemannian integration measure (in terms of the square-root of the
determinant of the Riemannian space-time metric). An important feature was the
requirement for global Weyl-scale invariance which subsequently underwent
dynamical spontaneous breaking \cite{Guendelman:1999qt,Ansari:2002kmo,Guendelman:1999tb,Guendelman:2002js}. The second action term
with the standard Riemannian integration measure  might also contain a
Weyl-scale symmetry preserving $R^2$-term \cite{Guendelman:2002js}.

The latter formalism yields various new interesting results
in all types of known generally covariant theories:

\begin{itemize}
\item
(i) $D=4$-dimensional models of gravity and matter fields containing 
the new measure of integration appear to be promising candidates for resolution 
of the dark energy and dark matter problems, the fifth force problem, 
and a natural mechanism for spontaneous breakdown of global Weyl-scale symmetry
\cite{Guendelman:1999qt,Ansari:2002kmo,delCampo:2010kf,delCampo:2011mq,delCampo:2015yfa,Guendelman:2013dka,Guendelman:2012gg,Guendelman:2014wqa}.
\item
(ii) Study of reparametrization invariant theories of extended objects 
(strings and branes) based on employing of a modified non-Riemannian 
world-sheet/world-volume integration measure \cite{Guendelman:2000mt,Guendelman:2000ph,Guendelman:2002fb} leads to dynamically 
induced variable string/brane tension and to string models of non-abelian 
confinement, interesting consequences from the modified measures spectrum \cite{Guendelman:2021zmf}, and construction of new braneworld scenarios \cite{Guendelman:2021bzk}. 
Recently \cite{Nishino:2014bfa} this formalism was generalized to
the case of string and brane models in curved supergravity background.
\item
(iii) Study in \cite{Guendelman:2013zya,Guendelman:2014lea} of modified supergravity models with an
alternative non-Riemannian volume form on the space-time manifold produces some
outstanding new features: 
(a) This new formalism applied to minimal $N=1$ supergravity 
naturally triggers the appearance of a dynamically generated cosmological constant
as an arbitrary integration constant, which signifies a new explicit
mechanism of spontaneous (dynamical) breaking of supersymmetry;
(b) Applying the same formalism to anti-de Sitter supergravity allows us to 
appropriately choose the above mentioned arbitrary 
integration constant so as to obtain simultaneously a very small effective
observable cosmological constant as well as a very large physical gravitino mass.
\end{itemize}
In this paper we will study a quintessential scenario where we will be driven from inflation to
a slowly accelerated phase describing our universe using a scale invariant two field model. 
Multifield inflation has been studied by several authors see for example \cite{Polarski:1994rz,Polarski:1992dq,Langlois:2008mn,Sa:2020fvn,Tellez-Tovar:2021mge}. In the context of modified measures formalism, the ratio of two measures can become an additional scalar field if we use the second order formalism \cite{Benisty:2018fja,Benisty:2019bmi}, in the present paper we will consider only the first order formulation however, and the measure field remain non dynamical, determined by a constraint and therefore they do not introduce new degrees of freedom.
Introducing two fields gives rise to very interesting  new possibilities. This is also the case when we consider multi field scale 
invariant inflationary models leading to DE/DM for the late universe, where interesting new features appear for both the inflationary phase and for the 
 DE/DM late universe phase, in particular we will see that the late universe acquires a fine structure with two possible vacuums for the late universe that can occur 
 at different times in the late evolution of the universe.

The plan of the present paper is as follows. In the next Section \ref{TMMT} we describe in
some detail the general formalism for the new class of gravity-matter
systems defined in terms of two independent non-Riemannian integration
measures. In Section \ref{flat-regions} we describe the properties of the three flat regions in the 
Einstein-frame effective scalar potential, one  corresponding to the evolution of
the early inflation and the other two for the late universe. We also present in this section the relevant solutions for the slow roll inflation.
In Section \ref{perturbations} we present a numerical analysis, for a reasonable choice of the 
parameters, of the resulting ratio of tensor-to-scalar 
perturbations and show that the present model conforms to the Planck Collaboration data.
In Section \ref{evolution} we study how the model can describe Dark Matter in a scale invariant fashion in the late Universe, what are the conditions for avoiding 5th force problem,
or what is equivalent for the dust Dark Matter to behave canonically. We find that in the two flat regions of the late Universe the Dark Energy and Dark Matter can acquire different parameters. In Section \ref{masses} we find  the different values of the particle masses  in the relevant vacuums of the two flat regions relevant to the late universe where the 5th force is eliminated, the dust is canonical etc. The dynamical connection between these two phases requires a non canonical dust and dark energy behavior transition, since particle masses have to change when transitioning between these two states. This has not been studied in full details yet.
We conclude in Section \ref{discuss} with some discussions. For simplicity 
we will use units where the Newton constant is taken as $G_{\rm Newton} = 1/16\pi$.

\section{Gravity-Matter Formalism With Two Independent Non-Riemannian Volume-Forms}
\label{TMMT}
In this section, we shall consider the following non-standard gravity-matter system with an action 
that involving two independent non-Riemannian integration
measure densities generalizing the model analyzed in \cite{Guendelman:2014waa}. In this form, the action is given by 
\be
\begin{split}
S = \int d^4 x\,\P_1 (A) \Bigl\lb R + L^{(1)} \Bigr\rb +  
\\\int d^4 x\,\P_2 (B) \Bigl\lb L^{(2)} + \eps R^2 + 
\frac{\P (H)}{\sqrt{-g}}\Bigr\rb \; ,    
\end{split}
\lab{TMMT}
\ee
where the following notations are used:

\begin{itemize}
\item
The quantities $\P_{1}(A)$ and $\P_2 (B)$ are two independent non-Riemannian volume-forms, 
\textsl{i.e.}, generally covariant integration measure densities on the underlying
space-time manifold and are given by:
\be
\P_1 (A) = \frac{1}{3!}\vareps^{\m\n\k\l} \pa_\m A_{\n\k\l},\quad
\P_2 (B) = \frac{1}{3!}\vareps^{\m\n\k\l} \pa_\m B_{\n\k\l} \; ,
\lab{Phi-1-2}
\ee
defined as a function of field-strengths of two auxiliary 3-index antisymmetric
tensor gauge fields \footnote{In general for the $D$ space-time dimensions one can always
represent a maximal rank antisymmetric gauge field $A_{\m_1\ldots\m_{D-1}}$
as a function of $D$ auxiliary scalar fields $\phi^i$ ($i=1,\ldots,D$) as:
$A_{\m_1\ldots\m_{D-1}} = \frac{1}{D}\vareps_{i i_1\ldots i_{D-1}} 
\phi^i \pa_{\m_1}\phi^{i_1}\ldots \pa_{\m_{D-1}}\phi^{i_{D-1}}$, so that its
(dual) field-strength
$\P(A) = \frac{1}{D!}\vareps_{i_1\ldots i_D} \vareps^{\m_1\ldots\m_D}
\pa_{\m_1}\phi^{i_1}\ldots \pa_{\m_D}\phi^{i_D}$.} 
The functions $\P_{1,2}$ take over the role of the standard Riemannian integration measure density defined as
$\sqrt{-g} \equiv \sqrt{-\det\Vert g_{\m\n}\Vert}$ and it is expressed  in terms of the space-time
metric $g_{\m\n}$.

\item
The functions $R = g^{\m\n} R_{\m\n}(\G)$ and $R_{\m\n}(\G)$ correspond to  the scalar curvature and the 
Ricci tensor in the first-order (Palatini) formalism, where the affine connection $\G^\m_{\n\l}$ is \textsl{a priori} independent of the metric $g_{\m\n}$. Also, we have added in the second action term  a $R^2$ gravity term (again in the Palatini form). We mention that $R+R^2$ gravity within the second order formalism (which was the first inflationary model) was originally analyzed in Ref.\cite{Starobinsky:1980te}.

\item
The quantities $L^{(1,2)}$ denote two different Lagrangians of two  scalar matter fields $\vp_1$ and $\vp_2$ in analogy to Ref.\cite{Guendelman:1999qt,Ansari:2002kmo}. These Lagrangians are defined as:
\br
L^{(1)} = -\h g^{\m\n} \pa_\m \vp_1 \pa_\n \vp_1 -\h g^{\m\n} \pa_\m \vp_2 \pa_\n \vp_2- V(\vp_1,\vp_2),
\lab{L-1} \\
L^{(2)} = U(\vp_1,\vp_2) ,
\lab{L-2}
\er
where the scalar  potential $V$ is given by
\be
V(\vp_1,\vp_2)=f_1\,e^{-\alpha_1\vp_1}+g_1e^{-\alpha_2\vp_2},
\ee
and the another scalar potential is defined as
\be
U(\vp_1,\vp_2)=f_2\,e^{-2\alpha_1\vp_1}+g_2\,e^{-2\alpha_2\vp_2},
\ee
where the quantities $f_1,f_2,g_1,g_2$,$\alpha_1$ and $\alpha_2$ are positive parameters. 
\item
The function $\P (H)$ denotes the dual field strength of a third auxiliary 3-index antisymmetric
tensor gauge field:
\be
\P (H) = \frac{1}{3!}\vareps^{\m\n\k\l} \pa_\m H_{\n\k\l} \; ,
\lab{Phi-H}
\ee
whose introduction is fundamental  for non-triviality of the model. 
\end{itemize}

We mention the scalar potentials $V$ and $U$  have been chosen in such a way that the  action given 
eq.\rf{TMMT} is invariant under global Weyl-scale transformations:
\begin{equation}
\begin{split}
g_{\m\n} \to \l g_{\m\n} \;\; ,\;\; \G^\m_{\n\l} \to \G^\m_{\n\l} \;\; ,\;\; 
 \vp_1 \to \vp_1+\frac{1}{\alpha_1}\ln\lambda\;,\\\vp_2 \to \vp_2+\frac{1}{\alpha_2}\ln\lambda,
A_{\m\n\k} \to \l A_{\m\n\k} \;\; ,\\\;\; B_{\m\n\k} \to \l^2 B_{\m\n\k}
\;\; ,\;\; H_{\m\n\k} \to H_{\m\n\k} \; .
\lab{scale-transf}
\end{split}
\end{equation}
 Note that this combination is invariant $\alpha_1\vp_1-\alpha_2\vp_2 \to \alpha_1\vp_1-\alpha_2\vp_2,$ from eq.\rf{scale-transf}.
Additionally, we observe that the requirement about the global Weyl-scale symmetry 
\rf{scale-transf} uniquely fixes the structure of the 
non-Riemannian-measure gravity-matter action  given by eq.\rf{TMMT}.



In the following we will use $\eps=0$ and this case the equations of motion resulting  from the
variation of \rf{TMMT} w.r.t. affine connection $\G^\m_{\n\l}$, are  
\be
\int d^4\,x\,\sqrt{-g} g^{\m\n} \Bigl(\frac{\P_1}{\sqrt{-g}} \Bigr) \(\nabla_\k \d\G^\k_{\m\n}
- \nabla_\m \d\G^\k_{\k\n}\) = 0 .
\lab{var-G}
\ee
Therefore,  
$\G^\m_{\n\l}$ corresponds to a Levi-Civita connection
\be
\G^\m_{\n\l} = \G^\m_{\n\l}({\bar g}) = 
\h {\bar g}^{\m\k}\(\pa_\n {\bar g}_{\l\k} + \pa_\l {\bar g}_{\n\k} 
- \pa_\k {\bar g}_{\n\l}\) \; ,
\lab{G-eq}
\ee
w.r.t. to the Weyl-rescaled metric ${\bar g}_{\m\n}$:
\be
{\bar g}_{\m\n} = \chi_1 \; g_{\m\n} \;\; ,\;\; \mbox{and}\,\,\,\,
\chi_1 \equiv \frac{\P_1 (A)}{\sqrt{-g}} \;.
\lab{bar-g}
\ee

Also, from the variation of the action \rf{TMMT} w.r.t. auxiliary tensor gauge fields
$A_{\m\n\l}$, $B_{\m\n\l}$ and $H_{\m\n\l}$ yields the equations, we have
\be
\begin{split}
\pa_\m \Bigl\lb R + L^{(1)} \Bigr\rb = 0, \quad
\pa_\m \Bigl\lb L^{(2)} +  \frac{\P (H)}{\sqrt{-g}}\Bigr\rb = 0 , \\ \pa_\m \Bigl(\frac{\P_2 (B)}{\sqrt{-g}}\Bigr) = 0 \; ,    
\end{split}
\lab{A-B-H-eqs}
\ee
whose solutions are given by
\be
\frac{\P_2 (B)}{\sqrt{-g}} \equiv \chi_2   ,\;\;\;
R + L^{(1)} = - M_1   , \;\;\;
L^{(2)} +  \frac{\P (H)}{\sqrt{-g}} = - M_2  .
\lab{integr-const}
\ee
Here the quantities $M_1$, $M_2$ and $\chi_2$ are integration constants. However, the constants  $M_1$ and $M_2$ are arbitrary  and dimensional and $\chi_2$
arbitrary and dimensionless. 

We mention that the  integration constant $\chi_2$ in eq.\rf{integr-const} preserves
global Weyl-scale invariance in eq.\rf{scale-transf}, whereas 
the appearance of the another integration constants $M_1,\, M_2$
signifies {\em dynamical spontaneous breakdown} of global Weyl-scale invariance 
under \rf{scale-transf} due to the scale non-invariant solutions 
in eq.\rf{integr-const}.


Also, varying the action \rf{TMMT} w.r.t. $g_{\m\n}$ and using relations \rf{integr-const} 
we have
\be
\chi_1 \Bigl\lb R_{\m\n} + \h\( g_{\m\n}L^{(1)} - T^{(1)}_{\m\n}\)\Bigr\rb = \frac{\chi_2}{2} \Bigl\lb T^{(2)}_{\m\n} + g_{\m\n} \; M_2
- 2 R\,R_{\m\n}\Bigr\rb\; ,
\lab{pre-einstein-eqs}
\ee
where $\chi_1$ and $\chi_2$ are defined in \rf{bar-g},
and the quantities $T^{(1,2)}_{\m\n}$ correspond to the energy-momentum tensors of the scalar
field Lagrangians with the standard definitions:
\be
T^{(1,2)}_{\m\n} = g_{\m\n} L^{(1,2)} - 2 \partder{}{g^{\m\n}} L^{(1,2)} \; .
\lab{EM-tensor}
\ee

Now, taking the trace of eq.\rf{pre-einstein-eqs} and using again second relation of 
eq.\rf{integr-const}, we find that the scale factor $\chi_1$ becomes
\be
\chi_1 = 2 \chi_2 \frac{T^{(2)}/4 + M_2}{L^{(1)} - T^{(1)}/2 - M_1} \; ,
\lab{chi-1}
\ee
where $T^{(1,2)} = g^{\m\n} T^{(1,2)}_{\m\n}$. 

Thus, considering the second relation of eq.\rf{integr-const} together with eq.\rf{pre-einstein-eqs}, we obtain  
 the Einstein-like form
\br
R_{\m\n} - \h g_{\m\n}R = \h g_{\m\n}\(L^{(1)} + M_1\)
+ \frac{1}{2}\(T^{(1)}_{\m\n} - g_{\m\n}L^{(1)}\)
\nonu \\
+ \frac{\chi_2}{2\chi_1 } \Bigl\lb T^{(2)}_{\m\n} + 
g_{\m\n}\,M_2 \Bigr\rb \; .
\lab{einstein-like-eqs}
\er

In this context, we can bring eqs.\rf{einstein-like-eqs} into the standard form of Einstein 
equations for the   metric ${\bar g}_{\m\n}$, 
\textsl{i.e.}, the Einstein-frame gravity equations  
\be
R_{\m\n}({\bar g}) - \h {\bar g}_{\m\n} R({\bar g}) = \h T^{\rm eff}_{\m\n},
\lab{eff-einstein-eqs}
\ee
in with the energy-momentum tensor  (analogously to \rf{EM-tensor}) 
\be
T^{\rm eff}_{\m\n} = g_{\m\n} L_{\rm eff} - 2 \partder{}{g^{\m\n}} L_{\rm eff} ,
\lab{EM-tensor-eff}
\ee
where the  effective Einstein-frame scalar field Lagrangian:
\be
L_{\rm eff} = \frac{1}{\chi_1}\Bigl\{ L^{(1)} + M_1 +
\frac{\chi_2}{\chi_1}\Bigl\lb L^{(2)} + M_2  
\Bigr\rb\Bigr\} \; ,
\lab{L-eff}
\ee
where  $L^{(1,2)}$ represent Lagrangian densities defined as
\be
L^{(1)} = \chi_1\, (X_1+X_2) - V \quad ,\quad L^{(2)} =  U \; ,
\lab{L-1-2-Omega}
\ee
with the potentials $V$ and $U$ as in relations \rf{L-1}-\rf{L-2}. Also, to write $L_{\rm eff}$ in terms of the Einstein-frame
metric ${\bar g}_{\m\n}$  we consider the short-hand notation for the
 kinetic terms 
\be
 X_1 \equiv - \h {\bar g}^{\m\n}\pa_\m \vp_1 \pa_\n \vp_1,\,\,\,\, X_2 \equiv - \h {\bar g}^{\m\n}\pa_\m \vp_2 \pa_\n \vp_2.
\lab{X-def}
\ee

By combining  eqs.\rf{chi-1} and \rf{EM-tensor-eff}, and taking into account \rf{L-1-2-Omega}, 
we obtain

\be
\chi_1 = \frac{2\chi_2\Bigl\lb U+M_2 \Bigr\rb}{(V-M_1)}
\, \; .
\lab{chi-Omega}
\ee
From eqs.\rf{chi-Omega} and \rf{L-eff}, we find at the
explicit form for the Einstein-frame scalar Lagrangian $L_{\rm eff}$
\be
L_{\rm eff} =  X_1+X_2  - U_{\rm eff}(\vp_1,\vp_2) \; ,
\lab{L-eff-final}
\ee
in which the effective scalar  potential $U_{\rm eff}(\vp_1,\vp_2)$ becomes
\be
\begin{split}
 U_{\rm eff} (\vp_1,\vp_2) \equiv 
\frac{(V - M_1)^2}{4\chi_2 \Bigl\lb U + M_2 \Bigr\rb}
\\= \frac{\(f_1 e^{-\a_1\vp_1}+g_1e^{-\a_2\vp_2}-M_1\)^2}{4\chi_2\,\Bigl\lb f_2 e^{-2\a_1\vp_1}+g_2 e^{-2\a_2\vp_2}+ M_2 \Bigr\rb} \; .
\end{split}
\lab{U-eff} 
\ee


We refer   that choosing  the ``wrong'' sign of the scalar 
potential $U$ (Eq.\rf{L-2})  in the initial non-Riemannian-measure
gravity-matter action \rf{TMMT} is necessary to end up with the right sign in the effective  potential \rf{U-eff} associated to scalar fields $\vp_1$ and $\vp_2$ in the physical Einstein-frame effective gravity-matter action given by eq.\rf{L-eff-final}. On the other hand, the overall sign of the other initial scalar potential $V$ (Eq.\rf{L-2}) is in fact irrelevant since changing its sign does not alter the positivity of effective  potential given by eq.\rf{U-eff}.

\section{Flat Regions of the Effective Scalar Potential}
\label{flat-regions}
\subsection{Flat Regions values}

\begin{figure}
 	\centering
\includegraphics[width=0.48\textwidth]{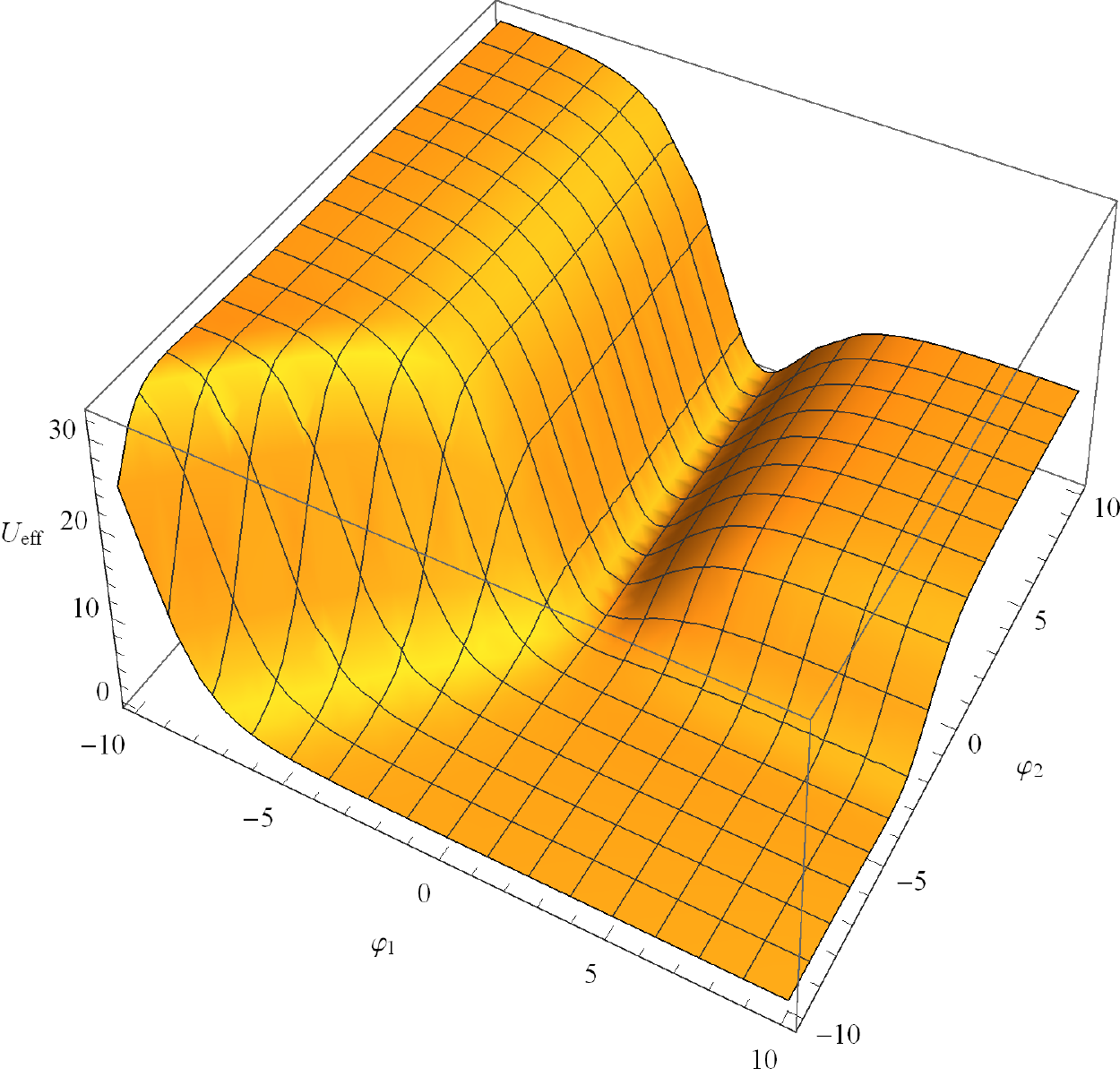}
\caption{\it{ {The effective potential with three flat regions. One flat region refers to the inflationary phase and the other region refers to dark energy. The third could be another early dark energy phase.Here we have used a positive value for $M_1$} }}
 	\label{fig:flatRegionsExample}
\end{figure}

We mention that the important feature of the effective potential $U_{\rm eff}$ (see eq.\rf{U-eff}) is the presence of three infinitely large
flat regions -- for large positive values of the  fields $\vp_1$ and $\vp_2$.
For large positive values of $\vp_1$ and $\vp_2$, we have for the effective potential reduces to

\br
U_{\rm eff}(\vp_1,\vp_2) \simeq U_{(++)} \equiv 
\frac{M_1^2}{4\chi_2\,M_2} \; .
\lab{U-minus} 
\er
 For the case in which we only  have  large negative $\vp_1$:
\br
U_{\rm eff}(\vp_1,\vp_2) \simeq U_{(\vp_1\to-\infty)} \equiv 
\frac{f_1^2}{4\chi_2 \,f_2} \; .
\lab{U-plus} 
\er
In the other flat region in which we only  have large negative $\vp_2$:
\br
U_{\rm eff}(\vp_1,\vp_2) \simeq U_{(\vp_2\to-\infty)} \equiv 
\frac{g_1^2}{4\chi_2 \,g_2} \; .
\lab{U-plus2} 
\er

 {Fig \ref{fig:flatRegionsExample} shows a qualitative example for the three fat regions.} The flat regions \rf{U-minus}, \rf{U-plus} and \rf{U-plus2} correspond to the evolution of the early and the late universe, respectively, provided we choose the ratio of the coupling constants in the original scalar potentials 
versus the ratio of the scale-symmetry breaking integration constants to obey:
\be
\frac{M_1^2}{M_2} \gg \frac{f_1^2}{f_2}, \,\, \mbox{and}\,\,\,\frac{M_1^2}{M_2} \gg \frac{g_1^2}{g_2},
\lab{early-vs-late}
\ee
which makes the vacuum energy density of the early universe $U_{(++)}$ much bigger
than that of the late universe.

On the other hand, from the cosmological perturbations together with the Planck data \cite{Planck:2014dmk,Planck:2013jfk,Planck:2018vyg,Planck:2018lbu,Planck:2019nip,BICEP:2021xfz}, we have that the first flat region of the effective potential is approximately  
\be
U_{(++)} \sim M_1^2/\chi_2M_2\sim 6\pi^2\,r\,\cP_S \sim 10^{-8}\,\;  ,
\lab{U-minus-magnitude}
\ee
(in units of $M_{Pl}^4$), where the $r$ denotes the tensor to scalar ratio and $\cP_S$ corresponds to the scalar power perturbation.
Let us recall that, since we are using units where $G_{\rm Newton} =
1/16\pi$, in the present case the Planck mass $M_{Pl}= \sqrt{1/8\pi G_{\rm Newton}} = \sqrt{2}$.

In order to study the dynamics of the universe, we consider that the metric corresponds to the  standard flat Friedmann-Lemaitre-Robertson-Walker
space-time metric given by: 
\be
ds^2 = - dt^2 + a^2(t) \Bigl\lb dr^2
+ r^2 (d\th^2 + \sin^2\th d\phi^2)\Bigr\rb,
\lab{FLRW}
\ee
where $a(t)$ denotes the scale factor. Thus,  
 the associated Friedmann equations
(recall the presently used units $G_{\rm Newton} = 1/16\pi$) result
\be
\frac{\addot}{a}= - \frac{1}{12} (\rho + 3p) \quad ,\quad
H^2  = \frac{1}{6}\,\,\rho \quad ,\;\; H\equiv \frac{\adot}{a} \; ,
\lab{friedman-eqs}
\ee
where $H$ is the Hubble parameter. Also,  the quantities $\rho$ and $p$ are defined as 
\br
\rho = \h  \vpdot_1^2 + \h  \vpdot_2^2+ U_{\rm eff}(\vp_1,\vp_2) \; ,
\lab{rho-def} \\
p = \h  \vpdot_1^2 + \h  \vpdot_2^2 - U_{\rm eff}(\vp_1,\vp_2),
\lab{p-def}
\er
and denote the total energy density and pressure of the scalar fields $\vp_1 = \vp_1 (t)$ and $\vp_2 = \vp_2 (t)$, respectively. 
In the following, we will consider that the dots indicate derivatives with respect to the time $t$.

In relation to  the scalar
 equations of motion for the scalar field $\vp_1$ and $\vp_2$, we have
\be
\vpddot_1 + 3 H \vpdot_1  + \partial U_{\rm eff}/\partial\vp_1 
 = 0 \; ,
\lab{vp-eqs-full}
\ee
and
\be
\vpddot_2 + 3 H \vpdot_2  + \partial U_{\rm eff}/\partial\vp_2 
 = 0 \; .
\lab{vp-eqs-full2}
\ee
 {From these equations it is useful to track the behavior of the solution for different values of the initial condition. From comparing the potential derivatives into zero $\partial U_{\text{eff}}/\partial \varphi_1  = \partial U_{\text{eff}}/\partial \varphi_2 = 0$ we get few points or paths. One path reads:}
\begin{equation}
f_1 e^{-\a_1\vp_1}+g_1e^{-\a_2\vp_2}=M_1,
\end{equation}
 {with $U_{\text{eff}} = 0$, which is possible if $M_1>0$ - the case we focus on. The path is a minimum from one side. The other points has infinite eigenvalues so we don't take them into account.}

 {Since the potential has three different flat regions that gives $\partial U_{\text{eff}}/\partial \varphi_1 = \partial U_{\text{eff}}/\partial \varphi_2 = 0$, the asymptotic behavior of the quintessential inflationary solution is quantifies by these areas. In early times the potential begins at $U_{(++)}$ and finishes at the lower value of the late dark energy.}

\subsection{Slow Roll approximation}

In the context of the slow roll inflation, we can introduce the standard ``slow-roll'' parameters \cite{Liddle:1992wi,Liddle:1993fq}:
\be
\vareps \equiv - \frac{\Hdot}{H^2}, \quad 
\eta_1 \equiv -\frac{\vpddot_1}{H\vpdot_1} \; ,\,\,\mbox{and}\,\,\,\,
\eta_2 \equiv -\frac{\vpddot_2}{H\vpdot_2}, 
\lab{SLR-def}
\ee
and under the slow-roll approximation $\vareps$, $\eta_1$ and $\eta_2\ll$ 1, thus  one ignores the terms with $\vpddot_{1,2}$, 
 so that the $\vp_1,\vp_2$-equations  of motion together with 
 the second Friedmann eq.\rf{friedman-eqs} simplify  to:
\be
\begin{split}
3H\vpdot_1 + \partial U_{\rm eff}/\partial\vp_1 \simeq 0,\,\,\,\,3H\vpdot_2 + \partial U_{\rm eff}/\partial\vp_2 \simeq 0  ,\\ H^2 \simeq\frac{1}{6} U_{\rm eff} \; .
\end{split}
\lab{slow-roll-eqs}
\ee

Since now the fields $\vp_1$ and $\vp_2$ evolve on the first flat region of $U_{\rm eff}$ for large positive values
\rf{U-minus}, we can consider that the effective  potential during inflationary scenario can be approximated to,  
\be
U_{eff}(\vp_1,\vp_2)\simeq\frac{M_1^2-2M_1(f_1e^{-\a_1\vp_1}+g_1e^{-\a_2\vp_2})}{4\chi_2M_2}.
\lab{Ueffapprox}
\ee
Here we have used the expansion of the effective potential given eq.\rf{U-eff} .

In the following we will introduce the number of $e-$folds $N$ defined as $N=\ln(a/a_f)$ where 
$a_f$ corresponds to the scale factor at the end of the inflation, that is, at the end of inflation $N=0$. Thus, 
 from eqs.\rf{slow-roll-eqs} and 
 \rf{Ueffapprox}
 can be rewritten as,
\be
\frac{d\vp_1}{dN}=\frac{6M_1\a_1f_1\,e^{-\a_1\vp_1}}{[M_2^2-2M_1(f_1e^{-\a_1\vp_1}+g_1e^{-\a_2\vp_2})]},
\ee
and
\be
\frac{d\vp_2}{dN}=\frac{6M_1\a_2g_1\,e^{-\a_2\vp_2}}{[M_2^2-2M_1(f_1e^{-\a_1\vp_1}+g_1e^{-\a_2\vp_2})]}.
\ee
Dividing these two equations we get a relation between the scalar fields $\vp_1$ and $\vp_2$ given by,
\be
e^{\a_1\vp_1}d\vp_1 =\frac{f_1\alpha_1}{g_2\alpha_2}
\,e^{\a_2\vp_2}d\vp_2.
\ee
Notice that the symmetry breaking constants  $M_1$ and $M_2$ dropped from this equation. The integration of this equation introduces a new constant of integration $C$
\be
e^{\a_1\vp_1}=\frac{f_1\a_1^2}{g_1\a_2^2}\,e^{\a_2\vp_2}+ C \lab{ref1f2}.
\ee
 In the following we will consider that
  the integration constant $C=0$.
 
Now, we can redefine  two new scalar fields $\phi_1$ and $\phi_2$, in terms of the scalar fields $\vp_1$ and $\vp_2$, such that
\be
\phi_1=\frac{\a_1\vp_1-\a_2\vp_2}{\sqrt{\a_1^2+\a_2^2}},\,\,\,\,\mbox{and}\,\,\,\,\phi_2=\frac{\a_2\vp_1+\a_1\vp_2}{\sqrt{\a_1^2+\a_2^2}}.
\lab{scalarfieldsphi}
\ee
Thus,  this transformation is orthogonal, $\dot{\phi}_1^2+\dot{\phi}_2^2=
\dot{\varphi}_1^2+\dot{\varphi}_2^2$,  where $\phi_1$ is invariant and $\phi_2$ transforms under a scale transformation.

 Notice that in this case, the scale
invariant combination $\a_1\vp_1-\a_2\vp_2$
gets determined, which corresponds  to fixing the scalar field $\phi_1$ defined in \rf{scalarfieldsphi}, this scalar field is scale invariant and is given by
\be
 \lab{slowrollfixedphi1}
\phi_1=\frac{1}{\sqrt{\a_1^2+\a_2^2}}\,\ln\left[\frac{f_1\a_1^2}{g_1\a_2^2}\right]=\mbox{constant}.
\ee
However, 
 the scalar field $\phi_2$ defined also equation in \rf{scalarfieldsphi},  evolves in time. This means that although we have broken the scale invariance, through the integration constants $M_1$ and $M_2$, some of the remaining equations recall such scale invariance. As we have noticed in particular, the integration constants $M_1$ and $M_2$ dropped from such equation. That is indeed the reason that
the equation that relates the two scalars retains the scale invariance, which is not true for other equations. 
We 
can now go back to the fields $\vp_1$ and $\vp_2$, in particular  we have that the relation between  the scalar field $\vp_2$ and the number of $e-$folds $N$ becomes
\be
\frac{A_2}{\a_2}e^{\a_2\vp_2}+A_3\vp_2=A_1N+cte,
\ee
and we can obtain $\vp_2=\vp(N)$ using the ProductLog function.
In mathematics, the product logarithm , also called the Omega function or  Lambert W function, is a multivalued function, namely the branches of the converse relation of the function $f(w) = we^w$. Using this definition, we find that the scalar field $\vp_2$ in terms of the number of $e$-folds results
\be
\begin{split}
\vp_2(N)=(A_1\,N+C_1)/A_3-\;
\\\a_2^{-1}\mbox{ProductLog}[(A_2/A_3)e^{\a_2(A_1\,N+C_1)/A_3}],\lab{FN1}    
\end{split}
\ee
where $C_1$ corresponds to another integration constant and the quantities $A_1$, $A_2$ and $A_3$ are defined as
$$
A_1=6M_1\a_2g_1,\,\,\,\,\,A_2=M_2^2,\,\,\,\,A_3=
-2M_1\left[\frac{g_1\a_2^2}{\a_1^2}+g_1\right].
$$
In order to obtain a real solution for the scalar field $\vp_2$ it is necessary that  the argument of the function ProductLog satisfies the condition in which the quantities 
$
(A_2/A_3)e^{\a_2(A_1\,N+C_1)/A_3}$
$>-e^{-1}$, see ref.\cite{doi:10.1080/00029890.1999.12005066}. 
 
 From eqs.\rf{ref1f2} and \rf{scalarfieldsphi} we find that the new scalar field $\phi_2$ can be written as
\be
\phi_2=\sqrt{\left[\left(\frac{\a_2}{\a_1}\right)^2+1\right]}\;\vp_2+C_2,
\ee
where $C_2$ is a constant defined as
$$
C_2=\frac{\a_2}{\a_1\sqrt{\a_1^2+\a_2^2}}\,\ln\left[\frac{f_1\a_1^2}{f_2\a_2^2}\right].
$$
Now, the effective potential associated to the new field $\phi_2$ becomes
\be
U_{eff}(\phi_2)\simeq\frac{M_1^2-2M_1\,g_1\left[\left(\frac{\a_2}{\a_1}\right)^2+1\right]e^{\frac{-\a_1\a_2(\phi_2-C_2)}{\sqrt{\a_1^2+\a_2^2}}}}{4\chi_2M_2}.\lab{P1}
\ee

In this way, the inflationary scenario reduces to a single field $\phi_2$, such that the new equations are $6H^2=\frac{\dot{\phi}_2^2}{2}+U_{eff}(\phi_2)$ and $\ddot{\phi}_2+3H\dot{\phi}_2+\partial U_{eff}(\phi_2)/\partial \phi_2$=0.

The new slow roll parameters
$\epsilon$ and $\eta$ associated to the scalar field $\phi_2$ are defined as in the standard case 
\be
\epsilon\simeq
\left(\frac{\partial U_{eff}/\partial \phi_2}{U_{eff}}\right)^2,\;\,\mbox{and}\,\,\,\,\eta\simeq2\left(\frac{\partial^2 U_{eff}/\partial \phi_2^2}{U_{eff}}\right)\label{P1}.
\ee
By considering the effective potential given by eq.\rf{P1} we obtain that the slow roll parameters result
\be
\begin{split}
\epsilon\simeq\left[\frac{4g_1^2\a_2^2(\a_1^2+\a_2^2)}{M_1^2\a_1^2}\right]\,\,e^{\frac{-2\a_1\a_2(\phi_2-C_2)}{\sqrt{\a_1^2+\a_2^2}}},\\\eta\simeq-\left[\frac{4g_1\a_2^2}{M_1}\right]\,\,e^{\frac{-\a_1\a_2(\phi_2-C_2)}{\sqrt{\a_1^2+\a_2^2}}}.\lab{ee2}    
\end{split}
\ee
Here we have considered that the effective potential $U_{eff}\sim M_1^2/(4\chi_2 M_2)$.

Additionally, we can obtain the value of $\phi_2$ at the end of the slow-roll regime $\phi_{2 end}$ and it  is determined
from the condition $\epsilon = 1$ which through \rf{ee2} becomes
\be
\phi_{2 end}=\frac{\sqrt{\a_1^2+\a_2^2}}{2\a_1\a_2}\,\ln\left[\frac{4g_1^2\a_2^2(\a_1^2+\a_2^2)}{M_1^2\a_1^2}\right]+C_2 .
\lab{vp-end}
\ee

Also, considering eq.\rf{FN1} we have that the value of scalar field $\phi_2$ at the end of inflation occurs when the number of $e$-folds $N=0$, and then 
\be
\begin{split}
\phi_{2 end}=C_2 + \sqrt{\left(\frac{\a_2^2}{\a_1^2}\right)^2+1}\\\left[\frac{C_1}{A_3}-\frac{1}{\a_2}
\mbox{ProducLog}[(A_2/A_3)e^{\a_2C_1/A_3}]\right].
\end{split}
\ee
Thus, from above equations, we obtain that the constant $C_1$ is given by
\be
C_1\approx \frac{A_3}{\a_2}\left[\frac{1}{2}\ln\left(\frac{4g_1^2\a_2^2(\a_1^2+\a_2^2)}{M_1^2\a_1^2}\right)+1\right].
\ee
Here we have considered that the term ProductLog  is a function that does not change very much and is of order one.

\section{Perturbations}
\label{perturbations}

In this section we will describe the scalar and tensor
perturbations during the inflationary stage for our model of the single field $\phi_2$. Following ref.\cite{Chen:2006nt,Bassett:2005xm}
the power spectrum of the scalar perturbation ${\cP_S}$
under the slow-roll approximation is defined as
\begin{equation}
{\cP_S} = \left(\frac{H^2}{2\pi\,\dot{\phi_2}}\right)^2 \simeq\left(\frac{1}{96\pi^2}\,\frac{U_{eff}^3}{(\partial U_{eff}/\partial\phi_2 )^2}\right) .\label{pec}
\end{equation}
The scalar spectral index $n_s$ is given by:
\be
n_s-1=\frac{d\ln \cP_S}{d\ln k}=
-6\epsilon+2\eta \; ,
\lab{ns}
\ee
where the slow roll parameters $\epsilon$ and $\eta$ are defined   by
eq.\rf{ee2}

On the other hand, it is well known that the generation of tensor perturbations in the scenario of 
inflation  would generate gravitational waves. In this context, the spectrum of the tensor perturbations
$\cP_T$ is defined as\cite{Chen:2006nt,Bassett:2005xm} 
\begin{equation}
\cP_T=\left(\frac{H}{\pi}\right)^2 \simeq \frac{U_{eff}}{6\pi^2}\; .
\label{Pt}
\end{equation}
Also, the tensor spectral index $n_T$ can be expressed in terms of
the slow parameter $\epsilon$ as $ n_T=\frac{d\ln \cP_T}{d\ln k}=-2\epsilon$.  

Additionally, an important observational quantity is the tensor-to-scalar ratio
$r=\frac{\cP_T}{\cP_S}$. We mention that these observational quantities should be evaluated when the cosmological
scale exits the horizon. In what follows the subscript $*$ is utilized  to indicate the epoch in which the cosmological
scale exits the horizon.

Considering the slow-roll approximation the power spectrum of the
scalar perturbation $\cP_S$  from eq.(\ref{pec}) can be written as 
\begin{equation}
\cP_S* \simeq\,\,k_1\,e^{\frac{2\alpha_1\a_2}{\sqrt{\a_1^2+\a_2^2}}(\phi_{2*}-C_2)},\lab{S1}
\end{equation}
where the constant $k_1$ is given by
$$
k_1=\left(\frac{1}{1536\pi^2}\right)\,\left(\frac{M_1^4}{\chi_2M_2g_1^2\,\a_2^2\,[(\a_2/\a_1)^2+1]} \right).
$$

From eq.\rf{ns} the scalar spectral index $n_s$, becomes
\be
\begin{split}
n_s*\simeq 1-\frac{8g_1\a_2^2}{M_1}\left[\frac{3g_1(\a_1^2+\a_2^2)}{M_1\a_1^2}e^{-\frac{\alpha_1\a_2}{\sqrt{\a_1^2+\a_2^2}}(\phi_{2*}-C_2)}+1\right]\\\times e^{-\frac{\alpha_1\a_2}{\sqrt{\a_1^2+\a_2^2}}(\phi_{2*}-C_2)} \; .
\lab{ns-1}
\end{split}
\ee
From eq.\rf{S1} we find that the quantity $\chi_2M_2g_1^4/M_1^6$ as a function of the power spectrum and the number of e- folds can be written as
\be
\begin{split}
 \frac{\chi_2M_2g_1^4}{M_1^6}=\left(\frac{g_1}{M_1}\right)^4\left(\frac{1}{4U_{(++)}}\right)\\=\left(\frac{1}{6144\pi^2}\right)\,\left(\frac{\a_1^4}{\a_2^2(\a_1^2+\a_2^2)^2\,\cP_S }\right)\,e^{\frac{6\a_2^2}{\sqrt{(\a_2^2/\a_1^2)+1}}N_*}\;.\lab{GGG}   
\end{split}
\ee
Also, considering eq.\rf{ns-1} we obtain that the ratio $g_1/M_1$ has four solutions and the real and positive solution is given by
\be
\begin{split}
\frac{g_1}{M_1}=\frac{\a_1^{3/2}}{\sqrt{12}\a_2^{1/2}(\a_1^2+\a_2^2)^{3/4}}\\\left[1+\sqrt{1+\frac{3}{2}\frac{(\a_1^2+\a_2^2)}{\a_1^2\a_2^2}(1-n_s)}\right]^{1/2}e^{\frac{3\a_1\a_2^2}{2\sqrt{\a_1^2+\a_2^2}}N_*}.\lab{gM}   
\end{split}
\ee
Additionally, we find that the tensor to scalar ratio $r$ as a function of the number of e-folds $N$ can be written as
\be
r(N=N_*)=r_*=\left(\frac{2\,g_1}{M_1}\right)^4\;\left[\frac{\a_2^4(\a_1^2+\a_2^2)^2}{\a_1^4}\right]\,e^{\frac{-6\a_1^2\a_2^2}{(\a_1^2+\a_2^2)}N_*},\lab{rrr}
\ee
here we have used  eqs.(\ref{Pt}) and  \rf{S1}.

By combining eqs.\rf{GGG} and \rf{rrr}, we find an upper bond for the parameter $\a_2$
given by 
\be
\a_2<r_*^{1/2}\,\left(\frac{6144\pi^2{\cP_S}_{*} }{2^6U_{(++)}}\right)^{1/2}.\lab{a222}
\ee
Also from eqs.\rf{GGG} and \rf{gM} we can obtain an equation that gives a relation between $\a_1$ and $\a_2$ given by 
\be
\left(\frac{3\gamma^2(\a_1^2+\a_2^2)^{1/2}}{\a_1}-1\right)^2=1+\frac{3}{2}(1-n_s)\left[\frac{(\a_1^2+\a_2^2)}{\a_1^2\a_2^2}\right],\lab{A1A2}
\ee
 {where $\gamma $ is defined as $\gamma=2[U_{(++)}/(1536\pi^2\cP_S)]^{1/4}$.}

In particular for the case in which the tensor to scalar ratio takes the value $r_{*}=0.036$, $\cP_{S*}\simeq 2.2\times 10^{-9}$ and the vacuum  energy $U_{(++)}\simeq6\pi^2\,r_{*}\,\cP_{S_*}\simeq 10^{-8}$ (see eq.\rf{U-minus-magnitude}) from ref.\cite{Planck:2014dmk,Planck:2013jfk,Planck:2018vyg,Planck:2018lbu,Planck:2019nip,BICEP:2021xfz,Campeti:2022vom}, we obtain from eq.\rf{a222} that the upper limit for the parameter $\a_2$ becomes 
$\a_2<2.74$. Now, using  this upper bound for $\alpha_2= 2.74$ and $n_{s_*}=0.967$, we find from eq.\rf{A1A2} that the real solution for $\a_{1}$ is given by $\a_{1}=0.24$. In the case in which $r_{*}=0.01$, we find that the upper bound for $\a_2\sim1.44$ and $\a_{1}\sim 0.07$. 

Additionally, in order to find a constraint for the ratio $g_1/M_1$, we can consider eq.\rf{GGG} (or \rf{gM}), obtaining that the the ratio $g_1/M_1$ for the special case in which $\a_1=0.24$, $\a_2=2.74$ and $N_*=60$ becomes $g_1/M_1\simeq 7\times 10^{25}$, and for $\a_1=0.07$, $\a_2=1.44$ we get  $g_1/M_1\simeq$ 4800.

 {Notice that the slow roll trajectory defined by \rf{slowrollfixedphi1} which for a given constant defines a straight line in the $(\varphi_1, \varphi_2)$ plane in the top vacuum
and for another constant defines another parallel line in the top vacuum. We can then choose the line we desire so as to fall in one of the two lower vacuua from the top vacuum.}



\section{Evolution to dark energy and dark matter}
\label{evolution}
In this section we will analyze the evolution of the dark energy and dark matter as a remnant of the early universe.
After the inflation period has ended there must be a period of particle creation that will produce dark matter as well as ordinary matter, this can be achieved in many different even in the case of one scalar field coupled to different measures \cite{Guendelman:2020zvt}. In this section we add now a dark matter particles contribution, defined in a scale invariant form by the matter action defined as
\begin{equation}
\label{particles}\\
S_{m}=\int( \Phi_1 +b_{m}e^{\kappa_1\p_2} \sqrt{-g})L_m d^{4}x ,
\end{equation}
 where $b_{m}$ is a constant that defines the strength to the coupling of $\p_2$ to $\sqrt{-g}$, coupling to $\Phi_2$ does not give a physically 
 different situation, since still $\Phi_2$ and $\sqrt{-g}$ are proportional. Also, the matter Lagrangian density  $L_m$ is given by
 \begin{equation}
L_m=-\sum_{i} m_{i}\int e^{\kappa_2\p_2}
\sqrt{g_{\alpha\beta}\frac{dx_i^{\alpha}}{d\lambda}\frac{dx_i^{\beta}}{d\lambda}}\,
\frac{\delta^4(x-x_i(\lambda))}{\sqrt{-g}}d\lambda ,\label{Lm}
\end{equation}
here the constants $\kappa_1$ and $\kappa_2$ satisfy the condition of scale invariance
and the quantity $m_i$ denotes the mass parameter of the ``{\it{i-th}}" particle. This invariace  determines the coupling constants to be equal to $\kappa_1 = -\frac{\alpha_1\a_2}{\sqrt{\a_1^2+\a_2^2}}$ and $\kappa_2 =-\frac{1}{2}\kappa_1$.

Under these conditions the presence of matter induces a potential for the scalar field $\p_2$ since there is a scalar field dependence $\p_2$ which  multiplies a ¨density
of matter¨contribution which is $\p_2$ independent. The scalar field $\p_2$ dependence is of the form,
\begin{equation}
(e^{-\frac{1}{2}\kappa_1\p_2}\Phi_1 + b_{m}e^{\frac{1}{2}\kappa_1\p_2} \sqrt{-g}).
\end{equation}

Such potential is extremized  by the condition
\begin{equation}\lab{condition}
\Phi_1 - b_{m}e^{\kappa_1\p_2} \sqrt{-g} = 0,
\end{equation}
interestingly enough the same condition eliminates all kind of non canonical anomalous effects, like the appearance of pressure in the contribution to the energy momentum from the particles,   { see section (\ref{beyondbackground})}.
Also the constraint equation that was used to determine the ratio of the measures $\Phi_1$ and $\sqrt{-g}$ becomes unaffected by the presence of the dust when the  condition above \rf{condition} is satisfied,  { see section (\ref{beyondbackground})}, so we can use equation \rf{chi-Omega} and in the late universe, neglecting $M_1$ and $M_2$, we obtain an equation that determines $\p_1$. Analogous  effects were recognized in a scale invariant two measure model of gravity, matter and one scalar field  in  \cite{Guendelman:2007ph} to obtain the avoidance of the Fifth Force Problem, which the $\p_2$, the ¨dilaton¨, could possibly cause, since it is a massless field. Here the the avoidance of the Fifth Force Problem is also achieved and we can arrange for this to happen when the scalar field $\p_1$ adjusts itself so as to satisfy the above equation. In this context, we find that the equation for $\phi_1$ is given by
 \be
 \begin{split}
 2\chi_2f_2e^{-\frac{\a_1^2}{\sqrt{\a_1^2+\a_2^2}}\p_1}+2\chi_2g_2e^{\frac{\a_1^2}{\sqrt{\a_1^2+\a_2^2}}\p_1}\\=b_mf_1+b_mg_1e^{\sqrt{\a_1^2+\a_2^2}\,\p_1}.\lab{ppp}     
 \end{split}
 \ee
 Thus,  eq.\rf{ppp} determines the value of $\phi_1$ to be a given constant solving this equation and then the velocity of the scalar field $\phi_1$ is zero i.e. $\dot{\p}_1=0$. In order to determine the value of the scalar field $\phi_1$ we consider $x=e^{\frac{\a_1^2\p_1}{\sqrt{\a_1^2+\a_2^2}}}$ then Eq.\rf{ppp} can be rewritten as
 \be
 2\chi_2g_2 x^2 - b_m g_1x^{\frac{2\a_1^2+\a_2^2}{\a_1^2}} - b_m f_1x + 2\chi_2 f_2    = 0,\lab{pp1}
 \ee
interestingly enough, the field  $\p_2$ drops from this equation. This is quite reasonable since  the field  $\p_2$ undergoes a shift under the scale transformation, so if  we were to determine  the field  $\p_2$ , that would correspond to a breaking of scale invariance, but now we are working in a phase with exact scale invariance, since we are neglecting the scale symmetry breaking constants $M_1$ and $M_2$. The field $\p_2$ is decoupled from matter, which is a consequence  of the elimination of the 5th force .

In order to obtain a solution for the scalar field $\phi_1$ from eq.\rf{ppp} or \rf{pp1} we consider that for very large value of $\phi_1$ or equivalently $x\rightarrow \infty$  the dominate terms of eq.\rf{pp1} are

\be  
2\chi_2g_2 x^2 - b_m g_1x^{\frac{2\a_1^2+\a_2^2}{\a_1^2}}\sim 0,\,\,\,\mbox{then}\,\,\,\,x\sim\left(\frac{2\chi_2g_2}{g_1 b_m}\right)^{(\a_1/\a_2)^2},\lab{x1}
\ee
where for consistency, we must choose the quantity $(\chi_2g_2/g_1b_m)\rightarrow \infty$.
Here the value of the scalar field $\phi_1$ at this point is
\be
\phi_{1_(+)}\sim\frac{\sqrt{\a_1^2+\a_2^2}}{\a_2^2}\,\ln\left[\frac{2\chi_2g_2}{f_1 b_m}\right].\lab{fp}
\ee

Now in the region in which the scalar field $\phi_1\rightarrow -\infty$ or $x\rightarrow 0$
we have that the dominant terms are 

\be  \lab{largenegativefield}
- b_m f_1x + 2\chi_2 f_2 \sim 0, \,\,\,\,\mbox{and}\,\,\,x\sim \left(\frac{2\chi_2 f_2}{f_1 b_m}\right)\rightarrow 0,
\ee
and the value of the scalar field at this point is
\be
\phi_{1_(-)}\sim\frac{\sqrt{\a_1^2+\a_2^2}}{\a_1^2}\,\ln\left[\frac{2\chi_2f_2}{f_1b_m}\right].\lab{fm}
\ee

In what follows of this section we study the dynamics of the dark energy and as defined before, with the equations for the ratio of the two measures obtained in the absence of dark matter \rf{chi-Omega} still being valid, so we can still consider the effective potential for the dark energy  by eq. \rf{U-eff} and the dark matter is described as a  dust since all non canonical effects disappear when $\Phi_1 - b_{m}e^{\kappa_1\p_2} \sqrt{-g} = 0$ is satisfied . When we also work in the very flat region, there is also no inconsistency with  $\p_1$ being a constant

The  flat-Friedmann equation for this stage is given by
\be
6H^2=\rho_{\vp_1,\vp_2}+\rho_{m},
\ee
where the energy density $\rho_{\vp_1,\vp_2}$ associated to the scalar fields $\vp_1$ and $\vp_2$ is 
\be
\rho_{\vp_1,\vp_2}=\frac{\dot{\vp}_1^2}{2}+\frac{\dot{\vp}_2^2}{2}+U_{eff}(\vp_1,\vp_2).
\ee
For the energy density of the dark matter $\rho_{m}$ we have
$$
\dot{\rho}_m+3H\rho_m=0,\,\,\,\mbox{then}\,\,\,
\rho_m(a)\propto\,\left(\frac{1}{a}\right)^3.
$$
From eq.\rf{U-eff} and considering  the region in which $f_1e^{-\a_1\vp_1}+g_1e^{-\a_2\vp_2}\gg M_1$ and $f_2e^{-2\a_1\vp_1}+g_2e^{-2\a_2\vp_2}\gg M_2$, the effective potential reduces to
\be
U_{eff}(\vp_1,\vp_2)=\frac{(f_1e^{-\a_1\,\vp_1}+g_1e^{-\a_2\vp_2})^2}{4\chi_2(f_2e^{-2\a_1\vp_1}+g_2e^{-2\a_2\vp_2})}.\lab{U4}
\ee
From eq.\rf{scalarfieldsphi} we have that the effective potential given by eq.\rf{U4} can be rewritten in term of the  single scalar field $\phi_1$ in which  
\be
U_{eff}(\phi_1)=\frac{(f_1e^{-\sqrt{\a_1^2+\a_2^2}\,\phi_1}+g_1)^2}{4\chi_2(f_2e^{-2\sqrt{\a_1^2+\a_2^2}\phi_1}+g_2)}.\lab{UUU}
\ee
 
As we have seen before the condition that the matter induced  potential of the scalar field $\phi_2$ is extremized requires the scalar field $\phi_1$ to be fixed at a very well specified point and now given the scalar field potential above, the eq. of motion of $\phi_1$ requires that this constant value be located at one of the two flat regions of the above potential at $\phi_{1_(+)}$ and $\phi_{1_(-)}$, see eqs.\rf{fp} and \rf{fm}. 
Thus, the energy density associated to the dark energy can be written as
\be
\rho_{\vp_1,\vp_2}=\rho_{\phi_1,\phi_2}=\frac{\dot{\phi}_1^2}{2}+\frac{\dot{\phi}_2^2}{2}+U_{eff}(\phi_1)=\frac{\dot{\phi}_2^2}{2}+U_{eff}(\phi_1),
\ee
where now the effective potential $U_{eff}$ depends only of the scalar field $\phi_1$. Also, as we have seen, the scalar field $\phi_1$ has been fixed to a constant because of the extremization of the $\phi_2$ matter induced potential, so we take  $\dot{\phi}_1=0$, see eq.\rf{ppp}. 

In order to study the evolution of the our model, we can choose the first flat region after inflation for the effective potential given by eq.\rf{UUU} assuming a very large scalar field $\phi_1$ given by $U_{eff +}=g_1^2/(4\chi_2 g_2)$, where the value of the scalar field $\phi_1$ is fixed in this flat region by eq.\rf{x1} $x\sim (2\chi_2g_2/g_1b_m)^{(\a_1/\a_2)^2}$=
$(g_1/[2b_m U_{eff +}])^{(\a_1/\a_2)^2}$ or equivalently $\phi_{1_(+)}$ defined by eq.\rf{fp}. For the second flat region after the inflation we can consider $\phi_1\rightarrow -\infty$ where the effective potential in this region is $U_{eff -}=(f_1^2/4\chi_2f_2)$. Here the value of the scalar field $\phi_1$ is fixed at $x\sim (2\chi_2f_2/f_1b_m)=(f_1/2b_m U_{eff -})$ or eq.\rf{fm}.

Additionally, 
 we can note that the scalar field $\phi_2$ corresponds to a massless field. In this way, the evolution of the scalar field $\phi_2$ as a function of the scale factor results 
\be
\ddot{\phi}_2+3H\dot{\phi}_2=0, \rightarrow\,\,\,\,\dot{\phi}_2=\frac{B_1}{a^3}=\dot{\phi}_{2+}\left(\frac{a_+}{a}\right)^3,\lab{f22}
\ee
where $B_1$ denotes an integration constant. By convenience $B_1=\dot{\phi}_{2+}\,a_+^3$, where $a_+$ and $\dot{\phi}_{2+}$ correspond to the scale factor and the velocity of the scalar field  in the first flat regime of the effective potential $U_{eff(+)}=g_1^2/(4\chi_2g_2)$.

The evolution of the scalar field $\phi_2$ as a function of the scale factor can be obtained considering that $\dot{\phi}_2=aH da/dt$, then eq.\rf{f22} can be rewritten as
\be
\frac{d\phi_2}{da}=\frac{B_1}{a^4\, H}, 
\lab{w1b}
\ee
and the Hubble parameter in terms of the scalar field is given by
\be
H=\frac{1}{\sqrt{6}}\left[\frac{B_1^2}{2a^6}+U_{eff(+)}+\frac{B_2}{a^3}\right]^{1/2},\,\,\,\mbox{with}\,\,\,\,B_2=\rho_{m_+}\,a_+^3,
\ee
where $\rho_m{_+}$ is the energy density associated to the dark matter in the first flat region of the effective potential $U_{eff(+)}$.
In particular, we have that in the first region the quantities $\rho_{m +}$ and $\dot{\phi}_{2+}$ become 
\be
\begin{split}
\rho_{m+}=6H_+^2\Omega_{m+},\,\,\,\,
\\
\,\,\,\,\,\dot{\phi}_{2+}=\left[2(6\,H_+^2\,\Omega_{{\phi_{+_1}},\phi_{+_2}}-U_{eff(+)})\right]^{1/2}.    
\end{split}
\ee
In this way, we find that the evolution of the scalar field as a function of the scale factor becomes
\be
\begin{split}
\phi_2(a)=\phi_{2_{+}}\\+\frac{2}{\sqrt{3}}\left
[\mbox{Arctanh}\left(\frac{B_1^2+B_2a_+^3}{B_1\sqrt{B_1^2+2a_+^3(B_2+U_{eff(+)}a_+^3)}}\right) \right] \\-\frac{2}{\sqrt{3}}\left[\mbox{Arctanh}\left(\frac{B_1^2+B_2\,a^3}{B_1\sqrt{B_1^2+2\,a^3(B_2+U_{eff(+)}\,a^3)}}\right) \right].  
\end{split}
\ee
Also, we can determine the equation of state (EoS) or EoS parameter $w$ associated to the scalar fields in terms of the scale factor given by
\be
w(a)=\frac{\frac{\dot{\phi}_2^2}{2U_{eff (+)}}-1}{\frac{\dot{\phi}_2^2}{2U_{eff (+)}}+1}=\frac{\left(\frac{2\chi_2g_2 B_1^2}{g_1^2\,}\right)\,a^{-6}-1}{\left(\frac{2\chi_2g_2 B_1^2}{g_1^2\,}\right)\,a^{-6}+1}.
\lab{w1a}
\ee
Additionally, the total EoS parameter  $w_T$ associated to dark matter and scalar fields becomes
\be
w_T=\frac{w}{(1+\rho_m/\rho_{\p_1,\p_2})},
\ee
and in terms of the scale factor the EoS parameter $w_T(a)$ is given by
\be
\begin{split}
w_T(a)=\left[\frac{\left(\frac{2\chi_2g_2 B_1^2}{g_1^2\,}\right)\,a^{-6}-1}{\left(\frac{2\chi_2g_2 B_1^2}{g_1^2\,}\right)\,a^{-6}+1}\right]\\\times\left(1+\frac{B_2\,a^{-3}}{(B_1^2/2)a^{-6}+(g_1^2/4\chi_2g_2)}\right)^{-1}.\lab{w332}    
\end{split}
\ee
We note that eq.\rf{w332} can be rewritten 
in terms of the density parameter $\Omega_{ +}$, by considering the Friedmann equation in which $1=\Omega_++\Omega_{m+}$, where $\Omega_+$ and $\Omega_{m+}$ denote the densities parameters of different components in the first flat region and then the EoS parameter becomes
\be
\begin{split}
w_T(a)=\left[\frac{(\Omega_{+}\,y_{+}-1)\tilde{a}^{-6}-1}
{(\Omega_{+}\,y_{+}-1)\tilde{a}^{-6}+1}\right]\\\times\left(1+\frac{y_+(1-\Omega_+)\tilde{a}^{-3}}{(\Omega_+\,y_+-1)\tilde{a}^{-6}+1}\right)^{-1},\lab{86}    
\end{split}
\ee
where the new scale factor $\tilde{a}$ is defined as $\tilde{a}=a/a_+$ and the quantity $y_+$ corresponds to the rate $y_+=6H_+^2/U_{eff(+)}$ and $H_+$ is the Hubble parameter in the first flat region. As the kinetic energy is defined as positive, then the condition for the quantity $y_+$ is $y_+>1/\Omega_+$. 

In fig.\rf{Fig1} we show the development of the total EoS parameter $w_T$ versus the scale factor $\tilde{a}=a/a_+$, in the first flat region of the effective potential $U_{eff (+)}$ for different values of the ratio $y_+=6H_+^2/U_{eff(+)}>1/\Omega_+$. We choose  that the value of the density parameter
of the dark energy in the flat region is $\Omega_+=0.85$. From the plot we observe that when we increase the ratio $y_+$ the total EoS parameter $w_T$ also increases. Also, we note that for values of the scale factor $a<a_+$, the universe does not present an accelerated phase, since the total EoS parameter $w_T$ approaches positive values. However, for values of $a\sim a_+$, we observe that the total EoS parameter  is $w_T<-0.3$ and the universe shows an accelerated expansion for  values of $y_+$ near to  $1/\Omega_+$.

\begin{figure}
	\centering
\includegraphics[width=0.48\textwidth]{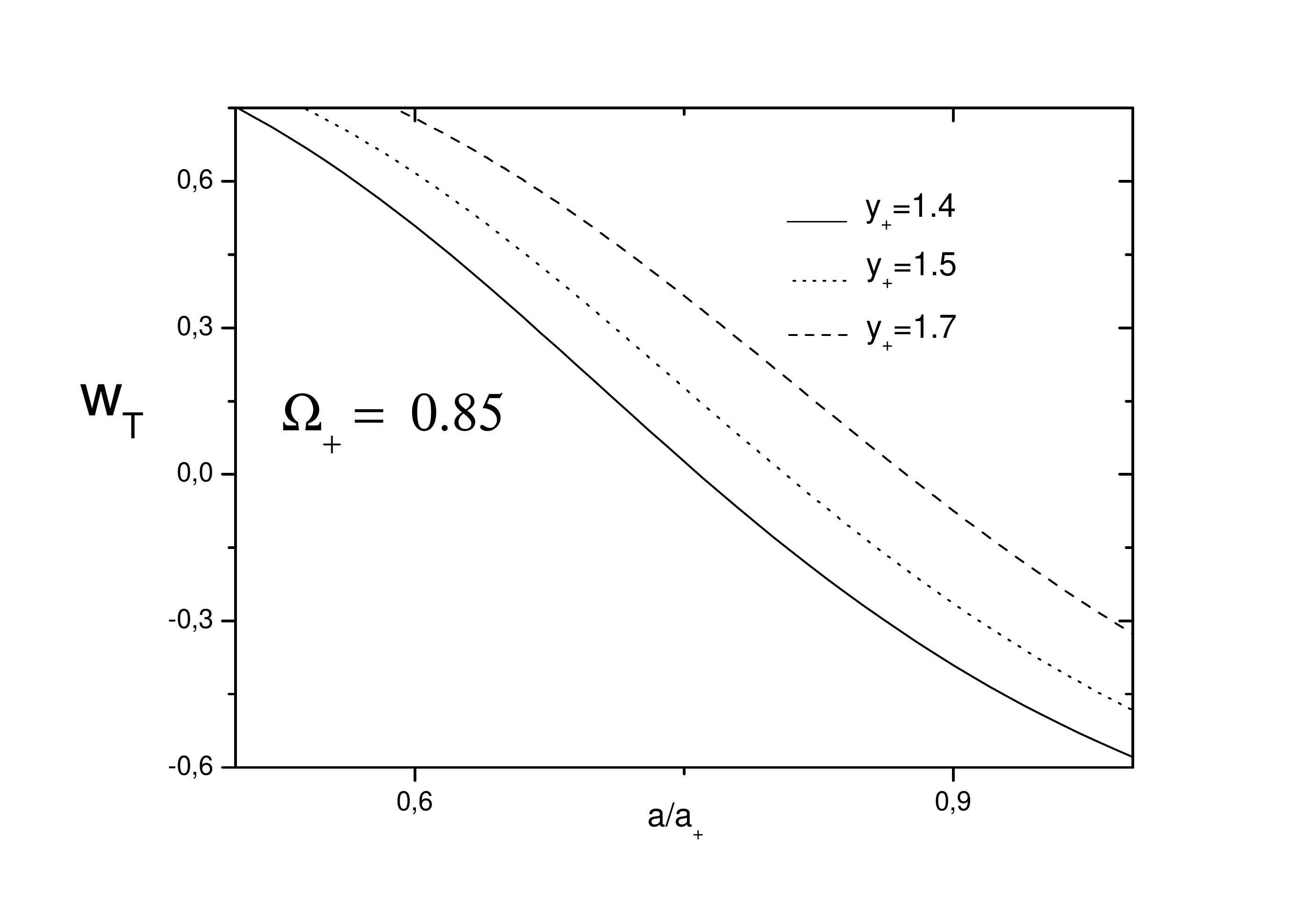}
\caption{\it{In this plot we show the evolution of the total EoS parameter  as a function of the scale factor $\tilde{a}=a/a_+$ in the first flat region of the potential $U_{eff (+)}=g_1^2/(4\chi_2g_2)$, for different values of the ratio $y_+=6H_+^2/U_{eff (+)}$, see eq.\rf{86}. In the first flat region we have used that the density parameter associated to the dark energy corresponds to   
 {$\Omega_+=0.85$, in order to satisfy the constraint from nucleosynthesis}. }}
\lab{Fig1}
\end{figure}

 { In addition, it is interesting to analyze the evolution of the barotropic parameter $w$ associated to  the dark energy. Following Refs.\cite{Caldwell:2005tm, Gupta:2011ip,Dutta:2011ik,delCampo:2010fg}, we can distinguish two categories from the behaviour of  ($dw/d\phi_2$); the  tracking freezing ($dw/d\phi_2<0$) and thawing ($dw/d\phi_2>0$) models. Thus, in our case for the first flat region and  considering that $dw/d\phi_2=(dw/d\tilde{a})(d\tilde{a}/d\phi_2)$  we find  
\be
\begin{split}
\frac{dw}{d\phi_2}=\frac{dw}{d\tilde{a}}\frac{d\tilde{a}}{d\phi_2}=\\-12\sqrt{12}\,\,\,\left(\frac{[\Omega_+\,y_+-1]}{1+y_+\Omega_+\tilde{a}^{-3}+[y_+-1-\Omega_+]\tilde{a}^{-6}}\right)^{1/2}\\\times\left(\frac{[\Omega_+y_+-1]}{[\Omega_+y_+-1+\tilde{a}^{6}]^2}\right)\,\tilde{a}\,<0.    
\end{split}
\ee
Here, we have used Eqs.\rf{w1b} and \rf{w1a}. In this form, we can infer that our model has a behaviour of freezing model, since $dw/d\phi_2$ is negative. This behaviour of the model occurs because that the ratio $dw/d\tilde{a}$ results negative i.e., $dw/d\tilde{a}<0$ and then we have a tracking freezing model.    
}
 {In this sense, we have that  in general for these tracking freezing models, are initially characterized by  $w>-1$ and $adw/da=dw/d\ln a<0$\cite{ZWS}. Thus, the tracker fields are characterized by  having  the attractor like solutions that converge to a common cosmic evolutionary track from different initial condition \cite{Steinhardt:1999nw,Zlatev:1998yg}. This suggests that the cosmology of the late time is independent of initial condition   due to 
 behavior of $w$ given by Eq.\rf{w1a}.  }

 {A new and important  constraint on the density energy associated to the dark energy during the radiation stage results from the nucleosynthesis.  It is well known that the quintessence scalar field modifies expansion of the universe at a given temperature
and in particular during the nucleosynthesis  where the temperature $T\sim $1 MeV, see \cite{Copeland:2006wr,Jaman:2022bho}.  Following Ref.\cite{Copeland:2006wr}, the energy density of the scalar field during this scenario can be constrained to $\Omega_{\phi}(T\sim 1$MeV)<7$\Delta N_{eff}/4/(10.75+7\Delta N_{eff}/4)$, where the value 10.75 corresponds to the effective number of standard model degrees of freedom and the quantity $\Delta N_{eff}$
denotes the additional relativistic degrees of freedom. In relation to the additional relativistic degrees of freedom in the literature it is considered as $\Delta N{eff}\simeq 1.5$ \cite{Kernan:1996yz} (see also Ref.\cite{Copi:1995cb} where $\Delta N_{eff}\simeq 0.9)$. Thus, considering  $\Delta N_{eff}\simeq 1.5$, then any quintessences models require to satisfy 
  $\Omega_{\phi}(T\sim 1$ MeV)<0.2 during the nucleosynthesis.
}

 { For our model we find that the density energy associated to dark energy $\Omega_{\phi_1,\phi_2}=\rho_{\phi_1,\phi_2}/6H^2$ can be written 
\be
\Omega_{\phi_1,\phi_2}(a)=\left(1+\frac{([\Omega_+\,y_+-1]\tilde{a}^{-6}+1)\tilde{a}^3}{y_+(1-\Omega_+)}\right)^{-1}.
\ee
Thus,   in order to satisfy the constraint imposed by the nucleosynthesis at the temperature $T\sim $1 MeV, we obtain the following bounds   
$$
\frac{4}{5}<\Omega_+<1, \,\,\,\,0<\tilde{a}_{T_* }^3<\frac{4-4\Omega_+}{\Omega_+},\,\,\,
$$
and 
\be
y_+>\frac{1-\tilde{a}_{T_*}^6}{\Omega_+(1+4\tilde{a}_{T_*}^3)-4\tilde{a}_{T_*}^3},
\ee
where  $\tilde{a}_{T_*}$ corresponds to the scale factor evaluated at the temperature $T_*= 1$MeV i.e.,  $\tilde{a}_{T_*}=a(T_*= $ 1 MeV)/$a_+$. Here we note that the   nucleosynthesis  epoch imposes a strong condition on the  density parameter  $\Omega_+$ and ratio $6H^2_+/U_{eff (+)}=y_+$ in the flat region.  }

 On the other hand, during the second flat regime associated to the effective potential $U_{eff (-)}$, the evolution of the scalar field $\phi_2$ as a function of the scale factor can be obtained considering as before that $\dot{\phi}_2=aH da/dt$, then eq.\rf{f22} can be rewritten as
\be
\frac{d\phi_2}{da}=\frac{\tilde{B}_1}{a^4\, H}, \,\,\,\,\mbox{where}\,\,\,\tilde{B_1}=\dot{\phi}_{02}a_0^3,
\lab{w2b}
\ee
where $\dot{\phi}_{02}$ and $a_0$ denote the velocity of the scalar field and the scale factor at the present epoch.

As before, the Hubble parameter in terms of the scale factor in this region is given by
\be
H=\frac{1}{\sqrt{6}}\left[\frac{\tilde{B}_1^2}{2a^6}+U_{eff(-)}+\frac{\tilde{B}_2}{a^3}\right]^{1/2},\,\,\,\mbox{with}\,\,\,\,\tilde{B}_2=\rho_{m 0}\,a_0^3,
\ee
where $U_{eff (-)}$ corresponds to the effective potential for very negative large scalar field $\phi_1$ and it is defined as $U_{eff(-)}=f_1^2/(4\chi_2f_2)$, from eq.\rf{UUU}.
Also, the value $\rho_{m 0}$ corresponds to the  dark energy of the  matter at the present epoch in which the scale factor $a=a_0=1$. 
From the Friedmann equation we have $1=\Omega_{\phi_1,\phi_2}+\Omega_m$, where $\Omega_{\phi_1,\phi_2}$ and $\Omega_m$ denote the densities parameters of the different components.  
In particular, from this equation we obtain that at present era the quantities $\rho_{m 0}$ and $\dot{\phi}_{0_2}$ become 
\be
\begin{split}
\rho_{m0}=6H_0^2\Omega_{m0},\\\dot{\phi}_{0_2}=\left[2(6\,H_0^2\,\Omega_{{\phi_{01}},\phi_{0_2}}-U_{eff(-)})\right]^{1/2},    
\end{split}
\ee
where from the observational data we have $\Omega_{m 0}\simeq 0.3$ and 
$\Omega_{{\phi_{01}},\phi_{02}}\simeq 0.7$.

Also, we obtain that the evolution of the scalar field as a function of the scale factor during this second scenario results
\be
\begin{split}
\phi_2(a)=\phi_{2{0}}\\+\frac{2}{\sqrt{3}}\left
[\mbox{Arctanh}\left(\frac{\tilde{B}_1^2+\tilde{B}_2a_0^3}{\tilde{B}_1\sqrt{\tilde{B}_1^2+2a_0^3(\tilde{B}_2+U_{eff(-)}a_0^3)}}\right) \right]\\-\frac{2}{\sqrt{3}}\left[\mbox{Arctanh}\left(\frac{\tilde{B}_1^2+\tilde{B}_2\,a^3}{\tilde{B}_1\sqrt{\tilde{B}_1^2+2\,a^3(\tilde{B}_2+U_{eff(-)}\,a^3)}}\right) \right].    
\end{split}
\ee
As before, we can determine the EoS parameter $w$ associated to the scalar fields in terms of the scale factor during this second flat region
\be
w(a)=\frac{\frac{\dot{\phi}_2^2}{2U_{eff -}}-1}{\frac{\dot{\phi}_2^2}{2U_{eff -}}+1}=\frac{\left(\frac{2\chi_2f_2 \tilde{B}_1^2}{f_1^2\,}\right)\,a^{-6}-1}{\left(\frac{2\chi_2f_2 \tilde{B}_1^2}{f_1^2\,}\right)\,a^{-6}+1}.\lab{w2a}
\ee
Also, we find that the total EoS parameter  $w_T=w_T(a)$ associated to dark matter and scalar fields during this scenario results

\be
\begin{split}
w_T(a)=\left[\frac{\left(\frac{2\chi_2f_2 \tilde{B}_1^2}{f_1^2\,}\right)\,a^{-6}-1}{\left(\frac{2\chi_2f_2 \tilde{B}_1^2}{f_1^2\,}\right)\,a^{-6}+1}\right]\\\times\left(1+\frac{\tilde{B}_2\,a^{-3}}{(\tilde{B}_1^2/2)a^{-6}+(f_1^2/4\chi_2f_2)}\right)^{-1}.\lab{w33}
\end{split}
\ee
Also, we can rewrite eq.\rf{w33} in terms of the  density parameter at present epoch  $\Omega_{\phi_{01,\phi_{02}}}=\Omega_-$ and then the total EoS parameter becomes
\be
\begin{split}
w_T(a)=\left[\frac{(\Omega_{-}\,y_{-}-1)a^{-6}-1}
{(\Omega_{-}\,y_{-}-1)a^{-6}+1}\right]\,\\\left(1+\frac{y_-(1-\Omega_-)a^{-3}}{(\Omega_-\,y_--1)\,a^{-6}+1}\right)^{-1} 
\end{split}
,\lab{93}
\ee
with the scale factor 
 $a/a_0=a$ and the quantity $y_-$ corresponds to the rate $y_-=6H_0^2/U_{eff(-)}$. As the kinetic energy is positive,  then we determine  that the condition for the parameter $y_->1/\Omega_-$.
In particular, we have that the density parameter at the present associated to dark energy  $\Omega_-\simeq0.7$, such that $y_->10/7$.

\begin{figure}
\centering
\includegraphics[width=0.48\textwidth]{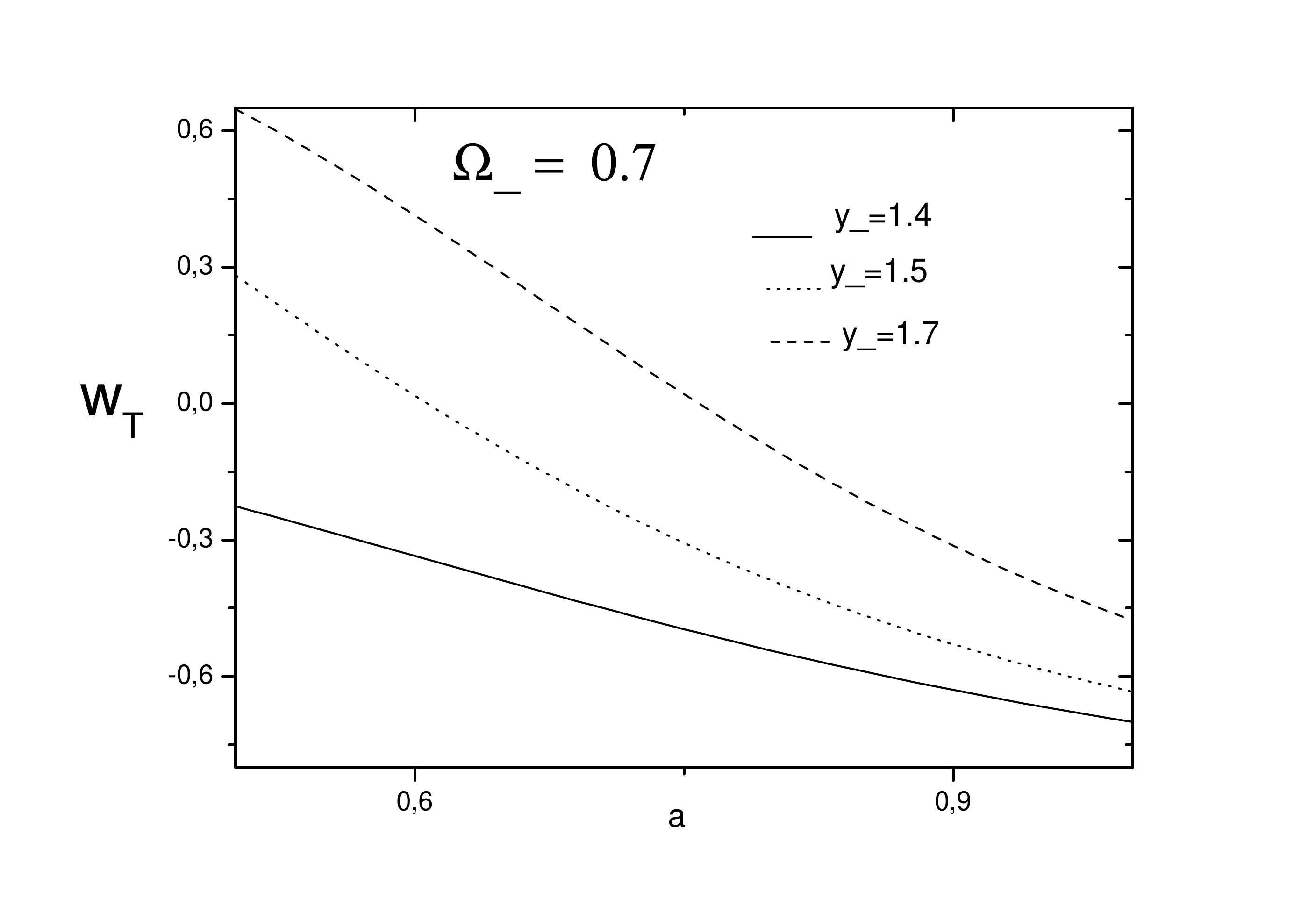}
\caption{\it{In this plot we show the evolution of the total EoS parameter $w_T$ as a function of the scale factor $a/a_0=a$ in the second flat region of the effective potential $U_{eff (-)}=f_1^2/(4\chi_2f_2)$. Here we have considered   different values of the ratio $y_-=6H_0^2/U_{eff (-)}$, in eq.\rf{93}. At the present time we have used that the density parameter associated to the dark energy is $\Omega_-=0.7$ and the scale factor $a_0=1$.}}
\lab{Fig2}
\end{figure}

In fig.\rf{Fig2} we show the evolution of the total EoS parameter $w_T$  versus the scale factor $a/a_0=a$ for different values of the ratio $y_-=6H_0^2/U_{eff (-)}>1/\Omega_-$. From the observational data we have considered that the density parameter  associated to the dark energy at the present era $\Omega_-=0.7$. As before in fig.\rf{Fig1}, from the plot we note that when we increase the ratio $y_-$, the total EoS parameter $w_T$ also grows. We observe that for values of the ratio $y_-\gg 1/\Omega_-$, the universe does not present an accelerated phase until now, since $w_T>-1/3$.


On other hand, 
we can obtain some estimates and constraints   on the parameter-space of the our model. For the second flat region of the effective potential $U_{eff (-)}$,  
 we can choose that the scales of the scale symmetry breaking integration constants
$f_1 \sim M^4_{EW}$ and $\chi_2f_2 \sim M^4_{Pl}$, where $M_{EW},\, M_{Pl}$ are
the electroweak and Plank scales, respectively.
In this case, we have 
a very small vacuum energy density $U_{(\p_1\to -\infty)}=U_{eff (-)}\sim f_1^2/\chi_2f_2$ given by
\be
U_{eff\,(\p_1\to -\infty)}=U_{eff (-)}\sim M^8_{EW}/M^4_{Pl} \sim 10^{-120} M^4_{Pl} \; ,
\lab{U-plus-magnitude}
\ee
where the mass $M_{EW}\sim 10^{-15} M_{Pl}$ and eq.\rf{U-plus-magnitude} corresponds to  the right order of magnitude for the present epoch's vacuum energy density, see  ref.\cite{Arkani-Hamed:2000ifx}. Thus, we can assume that the parameter $f_1\sim 10^{-60}$ (in units of Planck mass to the fourth power).

In order to transfer  the information of the inflationary stage to the present epoch, we can consider the constraints from inflationary scenario. In this context, we can utilize the constraint  from inflation for the  
ratio $g_1/M_1\sim 10^{26}$ for the special case in which $r_*=0.036$. In this way, we find that the effective potential in the fist flat region during the late universe $U_{eff (+)}$ can be written as 
\be
\begin{split}
U_{eff \,(\p_1\to \infty)}=U_{eff (+)}\simeq \frac{g_1^2}{4\chi_2g_2}\\\sim 10^{52}\,\frac{M_2\,U_{(++)}}{g_2}\sim 10^{44}\,\frac{M_2}{g_2}>U_{eff (-)}.    
\end{split}
\ee
Here we have considered that during inflation the energy density is $U_{(++)}\simeq 10^{-8}$. Also, as we have assumed that the effective potentials in the flat regions satisfied the condition $U_{eff (+)}>U_{eff (-)}\sim 10^{-120}$, then we find that lower bound for  the ratio $M_2/g_2$ becomes 
$
M_2/g_2> 10^{-164}$. In this form, we obtain that the ratio between the parameters associated to inflation ($M_1$ and $M_2$) and the first dark energy region ($g_1$ and $g_2$) results
\be
\frac{M_2}{M_1}>10^{-138}\,\frac{g_2}{g_1}.
\ee
Here we have used that the ratio $g_1/M_1\sim 10^{26}$.

\section{Dependence of the point particle masses on the scalar field $\phi_1$ and its consequences }
\label{masses}

One particular aspect that should be studied is the dependence of the point particle masses on the scalar field $\phi_1$ and its consequences.
We study this field dependence when the condition \rf{condition} is satisfied, which implies certain values of the scalar field $\phi_1$  are allowed. In this case we can solve the measure $\Phi_1$ using \rf{condition} and then considering the action in the Einstein frame. In such situation, a straightforward calculation shows that the masses of particles depend only on the scalar field $\phi_1$ in the following way,
\be \lab{massasfunctionofphi1}
\begin{split}
m_{ith-part} (\phi_1) = 2m_{i} b_m \\\times\sqrt{\frac{f_1e^{-\sqrt{\a_1^2+\a_2^2}\,\phi_1}+g_1}{2\chi_2(f_2e^{-2\sqrt{\a_1^2+\a_2^2}\phi_1}+g_2)}} e^{-\frac{\a_2^2\phi_1 }{2\sqrt{\a_1^2+\a_2^2}}}.    
\end{split}
\ee
As we can see from this equation, the particles in the solution with large $\phi_1$ , which correspond to the larger dark energy will have a much smaller mass
than the same particle when located at the vacuum with a much smaller value of $\phi_1$ . In a possible transition of these states, which will necessarily break condition  \rf{condition}  their DE and DM component will behave therefore in an opposite way after the process is completed and at the end point  \rf{condition} is restored again, so, as a result, when DE decreases, the DM component masses increase, of course, the DM component is still being diluted by the expansion of the universe, but enhanced by their increase in particle masses. As long as the particles remain in the states that satisfy \rf{condition}, the masses are fixed of course and the dust behaves canonically as described in the previous section. The discussion here concerns a transition between the two states that we have found that satisfy \rf{condition},
and the masses displayed by \rf{massasfunctionofphi1} concern masses only for such states. During the transition itself, the condition given by  eq.\rf{condition} must be violated, since this condition  allows only a discrete set of values, like those provided in \rf{x1}  and \rf{largenegativefield} only.

\section{Conditions for Canonical Dust Behavior Beyond The Background Case}
\label{beyondbackground}
 {In our previous considerations we have only considered cases where the scalar fields and the dust are distributed homogeneously in the Universe and we have also chosen the scalar field $ \phi_1$  by the observation that the presence of matter induces a potential for the scalar field $\p_2$ since there is a scalar field dependence $\p_2$ which  multiplies a ¨density of matter¨contribution which is $\p_2$ independent and the result of such minimization lead us to a value of $ \phi_1$ defined by eq.  \rf{condition}, which in turn lead us to a dust behavior for our model of point particles coupled in a scale invariant fashion. Here we will go a bit deeper, following the method studied in \cite{Guendelman:2007ph} for a single scalar field (for earlier treatments of the 5th force problems for more field theoretical models of matter rather than for point particle models of matter see \cite{Guendelman:2001xy,Guendelman:2002kw} ), and establish the more detailed conditions where this procedure can be more rigorously justified, we consider  $\chi_2 =1  $, since this constant can be reabsorbed into the definitions of particle densities, etc. in this case , For the purpose of this paper we restrict ourselves to a zero temperature gas of particles, i.e. we will assume that $d\vec{x}_i/d\lambda \equiv 0$  for all particles, which can be interpreted as the particles moving as co-moving, similar conclusions are easily derived without this assumption nevertheless. It is
convenient to proceed in the frame where $g_{0l}=0$, \, $l=1,2,3$. Then the particle density is defined by}
\begin{equation}
n(\vec{x})=\sum_{i}\frac{1}{\sqrt{-g_{(3)}}}\delta^{(3)}(\vec{x}-\vec{x}_i(\lambda)),
\label{n(x)}
\end{equation}
 {where $g_{(3)}=\det(g_{kl})$. We transform to the Einstein frame where this transformation causes the transformation of the particle density}
\begin{equation}
\bar{n}(\vec{x})=(\chi_1 )^{-3/2}\,n(\vec{x}). \label{nbar}
\end{equation}
 {The gravitational equations take the standard GR form}
\begin{equation}
G_{\mu\nu}(\bar{g}_{\alpha\beta})=\frac{\kappa}{2}T_{\mu\nu}^{eff},
 \label{gef}
 \end{equation}
 {where $G_{\mu\nu}(\bar{g}_{\alpha\beta})$ is the Einstein tensor in the Riemannian space-time with the metric $\bar{g}_{\mu\nu}$. The components of the effective energy-momentum tensor are as follows}
\begin{widetext}
\begin{equation}
T_{00}^{eff} = \left(\dot{\phi_1}^2- \bar{g}_{00}X_1\right) +  \left(\dot{\phi_2}^2- \bar{g}_{00}X_2\right) +\bar{g}_{00}\left[U_{eff}(\phi_1,\phi_2 ;\chi_1,M_1,M_2 ) +\frac{3\chi_1e^{-\frac{\kappa_1\p_2}{2}} +b_m e^{\frac{\kappa_1\p_2}{2}} }{2\sqrt{\chi_1
}}\, m\, \bar{n}\right], 
\label{T00}
\end{equation}
 {and}
\begin{equation}
T_{ij}^{eff} = 
\left(\phi_{1,k}\phi_{1,l}-\bar{g}_{kl}X_1\right) + \left(\phi_{2,k}\phi_{2,l}-\bar{g}_{kl}X_2\right) +\bar{g}_{kl}\left[U_{eff}(\phi_1,\phi_2 , \chi_1,M_1,M_2 )+\frac{\chi_1e^{-\frac{\kappa_1\p_2}{2}} -b_me^{\frac{\kappa_1\p_2}{2}}}{2\sqrt{\chi_1
}}\, m\, \bar{n}\right].
\label{Tkl}
\end{equation}
\end{widetext}
 {Here the following notations have been used:}
\begin{equation}
\begin{split}
X_{1}\equiv-\frac{1}{2}\bar{g}^{\alpha\beta}\phi_{1,\alpha}\phi_{1,\beta} \quad \text{and}\\
 X_{2}\equiv-\frac{1}{2}\bar{g}^{\alpha\beta}\phi_{2,\alpha}\phi_{2,\beta},\qquad
\end{split}
\end{equation}
 {and the function $U_{eff}$ is defined by}
\begin{equation}\label{energydensity}
U_{eff}(\phi_1,\phi_2 ;\chi_1)=
\frac{1}{\chi_1}\left[M_1-V\right]
+\frac{\chi_2}{\chi_1^2}(U+M_2),
\end{equation}
 {where $\chi_1$ has to be solved now for the case particles are present, which may differ somewhat with the solution in vacuum. The dilaton $\phi_2$ field equation is sourced by matter particles and in the Einstein frame is as follows,}
\begin{eqnarray}
&&\frac{1}{\sqrt{-\bar{g}}}\partial_{\mu}\left[\sqrt{-\bar{g}}\bar{g}^{\mu\nu}\partial_{\nu}\phi_2\right]
+ \frac{\partial U_{eff}}{\partial\phi_2}\nonumber\\
 &&=\kappa_1\,\frac{\chi_1 e^{-\frac{\kappa_1\p_2}{2}}
-b_m e^{\frac{\kappa_1\p_2}{2}}}{2\sqrt{\chi_1 } }\, m\,
\bar{n}.
 \label{phief}
\end{eqnarray}
 {In the above equations, the scalar field $\chi_1$  is determined as a function $\chi_1(\phi_1, \phi_2  , \bar{n})$ by means of the following constraint:}
\begin{eqnarray}
&&\frac{\chi_1\left(M_1 +V\right)-2\chi_2(U+M_2)}{(\chi_1)^2} =
\frac{\chi_1e^{-\frac{\kappa_1\p_2}{2}} -b_m e^{\frac{\kappa_1\p_2}{2}}}{2\sqrt{\chi_1 } }\, m\, \bar{n}.\nonumber\\
\label{constraint2}
\end{eqnarray}
 {In summary a ¨miracle ¨ takes place here, the same combination $\chi_1e^{-\frac{\kappa_1\p_2}{2}} -b_m e^{\frac{\kappa_1\p_2}{2}}$ appears in the right hand side of equations (\ref{constraint2}), (\ref{phief}) and in the anomalous pressure contribution produces by the dust displayed in (\ref{Tkl}). The vanishing of $\chi_1e^{-\frac{\kappa_1\p_2}{2}} -b_m e^{\frac{\kappa_1\p_2}{2}}$ was also obtained in our simplified considerations in eq. \rf{condition} from the condition of minimization of the matter induced potential for $\phi_2$, which (\ref{phief}) expresses in its full generality.}

 {The 5th force resolution for dense matter: Notice that in parallel to the idea of minimizing a matter induced potential, which gave us the vanishing of the right hand side of eq.  (\ref{phief}) we can look at eq.(\ref{constraint2}) as consisting of two parts, the right hand side can be compared with the energy density of  the scalar fields (\ref{energydensity}), so we can indeed say that this side is of the order of magnitude of this DE, but the other side on the other hand is proportional to the energy density of matter and for matter in ordinary state, which has energy density much larger than the vacuum energy of the universe, the only way to  have consistently is to have the coefficient   $\chi_1e^{-\frac{\kappa_1\p_2}{2}} -b_m e^{\frac{\kappa_1\p_2}{2}}$ that appears in the right hand side of equations (\ref{constraint2}) to be very close to zero. This coefficient represents the strength of the coupling of the scalar field $ \phi_2$ to matter.}

 {The next important issue to take notice is that once  $\chi_1e^{-\frac{\kappa_1\p_2}{2}} -b_m e^{\frac{\kappa_1\p_2}{2}}$ is taken to be zero, because this minimizes the matter induced potential for $\phi_2$, this leads us the vanishing of the right hand side of (\ref{constraint2}) and as a consequence to the same solution for $\chi_2$ as we obtained in vacuum, eq. \rf{chi-Omega}, which means that we can use the expressions for the effective potential in vacuum, now in the presence of dust. The dust is now totally canonical, as we have assumed in sections above, provided  $\chi_1e^{-\frac{\kappa_1\p_2}{2}} -b_m e^{\frac{\kappa_1\p_2}{2}} =0 $, which determine spacial values for $\phi_1$ in each of the flat regions of the effective potential as we have seen.}

 {Finally, the resulting effective potential in these flat regions is absolutely independent of $\phi_2$ as we have seen, so $\frac{\partial U_{eff}}{\partial\phi_2} = 0$ 
and furthermore there is no source since $\chi_1e^{-\frac{\kappa_1\p_2}{2}} -b_m e^{\frac{\kappa_1\p_2}{2}}$ is taken to be zero, so that indeed,  (\ref{phief}) implies then that  $\frac{1}{\sqrt{-\bar{g}}}\partial_{\mu}\left[\sqrt{-\bar{g}}\bar{g}^{\mu\nu}\partial_{\nu}\phi_2\right] =0$ as we have assumed.}

 {Let us analyze consequences of this wonderful coincidence in the case when the matter energy density (modeled by dust) is much larger than the dilaton contribution to the dark energy density in the space region occupied by this matter. Evidently this is the condition under which all tests of Einstein's GR, including the question of the fifth force, are fulfilled. if the dust is in the normal conditions there is a possibility to provide the desirable feature of the dust in GR: it must be pressureless. This is realized provided that in normal  conditions (n.c.) the following equality holds with extremely high accuracy:}
\begin{equation} \lab{decoupling-cond}
 \chi_1^{(n.c.)}\approx b_m  e^{\kappa_1\p_2}. 
\end{equation}
 {Remind that we have assumed  $b_m >0$
 Inserting the above equation in the last term of Eq. (\ref{T00}) we obtain the effective dust energy density in normal conditions, where the dependence on $\p_2$ has disappeared, as it should be for acceptable resolution of the 5th force problem.}
\begin{equation}
 \rho_m^{(n.c.)}=2\sqrt{b_m} \, m\tilde{n}.
\label{rho-m-n.c.}
\end{equation}
 {When we get only a slight deviation of from $\chi_1$ from $b_m e^{\kappa_1\p_2}$, when the matter energy density is many orders of magnitude larger than the dilaton contribution to the dark energy density, we obtain an effective 5th force  coupling  $f$. For this look at the $\phi$-equation in the form (\ref{phief}) and estimate the Yukawa type coupling constant in the r.h.s. of this equation. In fact, using the constraint (\ref{constraint2}) and representing the particle density in the form $\tilde{n}\approx N/\upsilon$ where $N$ is the number of
particles in a volume $\upsilon$, one can make the following estimation for the effective dilaton to matter coupling "constant" $f$ defined by the Yukawa type interaction term $f\bar{n}\phi$ (if we were to invent an effective action whose variation with respect to $\phi$ would result in Eq. (\ref{phief})):}
\begin{equation}
f \equiv\kappa_1 \,\frac{\chi_1 e^{-\kappa_1\p_2/2} -b_me^{\kappa_1\p_2/2}}{2\sqrt{\chi_1}}\approx
\kappa_1\,\frac{\rho_{vac}}{\tilde{n}} \approx
\kappa_1\frac{\rho_{vac}\upsilon}{N}.\label{Archimed}
\end{equation}
 {If we consider that $\kappa_1$ is a number divided by the Planck Mass, then $f$ becomes less than the ratio of the "mass of the vacuum" in the volume occupied by the matter to the Planck mass. The model yields this kind  of "Archimedes law" without any especial (intended for this) choice of the underlying action and without fine tuning of the parameters. The model not only explains why all attempts to discover a scalar force correction to Newtonian gravity were unsuccessful so far but also predicts that in the near future there is no chance to detect such corrections  in the astronomical measurements as well as in the specially designed  fifth force experiments on intermediate, short (like millimeter) and even ultrashort (a few nanometer) ranges. This prediction is alternative to predictions of other known models.}

 {Finally, we want to point out fundamental differences of our solution of the fifth force force problem to the Chameleon approach. The important point to make is that we are talking of totally different mechanisms, in the Chameleon model, the proposed quintessential scalar, the Chameleon field has a mass in vacuum which is very small, of the order of the Hubble parameter for example (or in any case very very small). The Chameleon scalar however becomes massive in presence of dense matter,  in compact objects, like Earth, a typical number for this mass has been cited, $ m^{-1}\sim 60_{mm}$ \cite{Khoury:2003aq}. This is why a quanta of this scalar field  can penetrate only into a thin shell of the body in the depth about 60micrometer, and the fifth force acts only on the thin shell. This is a way the Chameleon model is argued to explain the smallness of the fifth force. In our case there is no mass generation whatsoever since for our dilaton field, what happens here is the vanishing of the effective coupling constant between the dilaton field and the dense matter, while the dilaton keeps is mass zero or very close to zero. The elimination of interaction between our dilaton field and dense matter is total and absolute, in comparison, a Chameleon wave can  suffer a total reflection from a dense matter region, in such a situation it will not be a total elimination of the fifth force, but it may be hard indeed to prepare such an experiment. The elimination of the fifth force in the Chameleon model is argued to exist because in a spherically symmetric static configuration of a macroscopic object only a very small shell of the object can be a source of the Chameleon scalar, while in our case there would be no source for the scalar, not even the edge or surface of the dense object or at any place of the dense object. Higher-order theories of gravity also have been also studied in connection of fifth force suppression and have been shown to produce an explicit realization of the Chameleon scenario from first principles \cite{Brax:2004qh,Brax:2004ym,Brax:2004px,Capozziello:2007eu,Capozziello:2012ie}. }


\section{Discussion}
\label{discuss}

In the present paper we have constructed a new kind of gravity-matter theory
defined in terms of two different non-Riemannian volume-forms (generally
covariant integration measure densities) on the space-time manifold. We also introduced two scalar fields in a scale
invariant way. The integration of the equations of motion of the degrees of freedom that define the measures provides the 
constants of integration $M_1$ and $M_2$ which provide us with the spontaneous breaking of scale invariance. In the early universe inflation
 $M_1$ and $M_2$ play an important role, determining the scale of the inflationary energy density and defining the slow roll features in the inflationary epoch.
 In the slow roll solutions we  have studied one linear combination of the scalar fields $\vp_1$ and $\vp_2$, which we have called $\phi_1$, that remains constant during the inflationary phase. This combination is invariant under scale transformations, see eq.\rf{scale-transf}.
 
 The dynamics of inflation reduces to that of only one scalar field ($\phi_2$), but the full range of parameters obtained from the original two scalar field couplings, which have different couplings to the different measures plays a fundamental role. The allowed parameters range allowed from observations is studied.
 This study of allowed parameter ranges in inflation imposes constraints  on the parameter ranges in the late universe, where DM in addition to DE has to be considered.
  {We have recognized also under which conditions we will fall from inflation to one of the two possible low vacuum energy DE states, since the slow roll trajectory defined by \rf{slowrollfixedphi1} which for a given constant defines a straight line in the $(\varphi_1, \varphi_2)$ plane
and for another constant defines another parallel line. We can then choose the line we desire (corresponding to a choice of initial conditions) to fall in one of the two lower vacuua from the top vacuum.}
 The DE/DM sector in the late universe is determined by a dynamics where the 
constants of integration $M_1$ and $M_2$,  which provide us with the spontaneous breaking of scale invariance,  can be ignored. In this situation the scalar field potential that depends only on $\phi_1$ allows two different flat regions for possible dark energy sectors. In each of these sectors there are particular values of $\phi_1$ where the matter induces potential for $\phi_2$ is stabilized. At those points the matter behaves canonically, i.e. the dust does not produce pressure, etc.,
but in these two different regions the point particle masses are different. The scalar field $\phi_2$ remains a massless field in the two flat regions.   {Notice that in the present treatment, DE and DM are not unified, although, there is the possibility of unifying also  DE and DM \cite{Guendelman:2012gg}, such unification has not been studied here in the context of the quintessential inflation and transition to a slowly accelerated phase, but it may be a possible generalization in a future research. What has been done here however is to introduce the dark matter in a completely scale invariant form and we have shown explicitly the conditions under which this DM behaves as canonical dust, which is not trivial because of the couplings to the scalar field $\phi_2$, which is massless and when possible 5th force effects from this massless field disappear in each of the two flat regions that can describe DE.}


The above implies that the two flat regions at the values of $\phi_1$ where the matter behaves canonically  contain the following three elements: a constant DE, a DM component and a massless scalar field, these components differ in the two different regions. For these regions in the later universe, we have chosen the first flat region for the effective potential given by $U_{eff (+)}$, that corresponds to large scalar field $\phi_1$, i.e., $U_{eff (\phi_1\to \infty)}=U_{eff (+)}$. For the vacuum energy density at the present epoch, we have chosen the effective potential   $U_{eff (\phi_1\to -\infty)}=U_{eff (-)}$, such that $U_{eff(+)}>U_{eff(-)}$. For this scenario in which $U_{eff(+)}>U_{eff(-)}$ is reasonable to consider that the scalar field $\phi_1$  that remains fixed  is $\phi_{1(+)}$, see eq.\rf{fp}.  
Also, for  both regions, we have found analytically the evolution of the scalar field as a function of the scale factor   and also the total EoS parameter in terms of the scale factor i.e., $w_T=w_T(a)$. From the total EoS parameter, we have observed that for values of the ratio $y_{\pm}$ much bigger than the density parameter associated to dark energy $\Omega_{\pm}$, the universe does not present an accelerated phase and then the model does not work. However, for values of $y_{\pm}\sim\Omega_{\pm}$, we have found that in both scenarios in which the effective potential 
corresponds to a flat region, the universe presents an accelerated expansion, since the total EoS parameter $w_T<-0.3$.  { Also, we have found from the barotropic parameter associated to the dark energy that our model corresponds to a  tracking freezing model and this indicates that the cosmology of the late time is independent of the initial condition, once that one of the different  slow roll trajectories is defined so to fall into one of the two lower vacuua from the top vacuum.   
To complement the constraints on the parameter-space during  epoch dominated by  dark energy, we have considered the constraint from the nucleosynthesis,    which imposes a strong condition on the density parameter and the ratio between the Hubble parameter and the effective potential in the flat region.}

Another possibility that could occur is  the inverse situation in which $U_{eff(-)}>U_{eff(+)}$ and the scalar field $\phi_1$ in this  scenario should be  $\phi_{1(-)}$.
Also, an interesting situation that could take place  is that the second flat region of the effective potential associated to the dark energy will be in the future and has not yet been part of the history of the universe. Also, we have found from Planck data the different constraints on the parameters associated to our model during the inflationary stage and these values are considered to obtain constraints relevant to the DE/DM epoch. The dynamical connection between these two regions of the late universe may provide interesting clues concerning cosmological puzzles like the $H_0$ tension \cite{Kamionkowski:2014zda,Poulin:2018dzj,Poulin:2018cxd}.

\begin{acknowledgements}
E.G. want to thank the Universidad Cat\'olica de Valpara\'{\i}so, Chile,  for hospitality during this collaboration,  FQXi  for great financial support for work on this project at BASIC in Ocean Heights, Stella Maris, Long Island,  Bahamas  and  CA16104 - Gravitational waves, black holes and fundamental physics and  CA18108 - Quantum gravity phenomenology in the multi-messenger approach
for additional financial support and we want to thank the Miami2021 conference for inviting us to present our results through a talk, see \href{https://cgc.physics.miami.edu/Miami2021/Guendelman2.pdf}{here}. D.B gratefully acknowledge the supports of the Blavatnik and the Rothschild fellowships.
\end{acknowledgements}

\bibliographystyle{apsrev4-1}
\bibliography{ref}

\end{document}